\documentstyle[epsfig,11pt]{article}

\newcommand{\be}{\begin{equation}}
\newcommand{\ee}{\end{equation}}
\newcommand{\ben}{\begin{eqnarray}}
\newcommand{\een}{\end{eqnarray}}
\newcommand{\bc}{\begin{center}}
\newcommand{\ec}{\end{center}}

\begin{document}

\title{Statistical mechanics and the description of the early
universe\\ II. Principle of detailed balance and primordial $^4$He
formation}

\author{M. E. Pessah$^{a,b}$ and Diego F. Torres$^{b}$
\\
{\small $^a$Facultad de Ciencias Astron\'omicas y Geof\'{\i}sicas,
UNLP, Paseo del Bosque s/n, 1900, La Plata, Argentina}
 \\ {\small $^b$Instituto Argentino de
Radioastronom\'{\i}a, C.C.5, 1894 Villa Elisa, Buenos Aires,
Argentina} }

\date{\today }

\maketitle

\begin{abstract}
If the universe is slightly non-extensive, and the distribution
functions are not exactly given by those of Boltzmann-Gibbs, the
primordial production of light elements will be non-trivially
modified. In particular, the principle of detailed balance (PDB),
of fundamental importance in the standard analytical analysis, is
no longer valid, and a non-extensive correction appears. This
correction is computed and its influence is studied and compared
with previous works, where, even when the universe was considered
as an slightly non-extensive system, the PDB was assumed valid. We
analytically track the formation of Helium and Deuterium, and
study the kind of deviation one could expect from the standard
regime.  The correction to the capture time, the moment in which
Deuterium can no longer be substantially photo-disintegrated, is
also presented. This allows us to take into account the process of
the free decay of neutrons, which was absent in all previous
treatments of the topic. We show that even when considering a
first (linear) order correction in the quantum distribution
functions, the final output on the primordial nucleosynthesis
yields cannot be reduced to a linear correction in the abundances.
We finally obtain new bounds upon the non-extensive parameter,
both comparing the range of physical viability of the theory, and
using the latest observational data.
\end{abstract}

\bigskip


\newpage



\section{Introduction}

Primordial nucleosynthesis provides an interesting testing arena
where to analyze the viability of physical theories, particularly,
of the statistical description. It is in this epoch where the
earliest bounds upon a given theory with cosmological influence
can be imposed. Thermal processes (see Ref. \cite{PESSAH},
hereafter referred as Paper I) are non-trivially modified by a
non-extensive correction to quantum distribution functions. Then,
different abundances of light elements are a possible outcome.

Some of the predictions for primordial nucleosynthesis in a
non-extensive setting have been analyzed before by some of us,
using the asymptotic approach of the quantum distribution
functions, see Refs. \cite{PRL,PHYSICA}. Here, instead, we shall
consistently continue within the formalism given in Paper I. Since
this approach is simpler, we shall obtain analytical results far
beyond the point where previous works have reached, see for
instance Ref. \cite{PHYSICA,TIR}. Together with Paper I, we shall
then provide a complete history of the early universe, accordingly
modified taking into account a non-extensive setting. In this
paper, we shall focus on the study of the changes that
non-extensive statistics introduces in the principle of detailed
balance, for which we provide a detailed analysis, both numerical
and analytical. We shall then enhance the study presented in
\cite{PHYSICA}, by framing it in a larger picture which
encompasses an smaller number of approximations and a larger
number of predictions.

Primordial nucleosynthesis was recently used as well to
demonstrate that macroscopic samples of neutrinos in thermal
equilibrium are indeed distributed according to Fermi-Dirac
statistics \cite{CU}. These latter authors considered that
neutrinos were distributed by a Bose-Einstein statistics, and
established bounds (not very restrictive though) upon this
unexpected possibility. It is interesting to compare with our
case: we assume that neutrinos are indeed 1/2 particles, as a
large amount of data coming from particles accelerators show, but
that even when continuing being fermions, and fulfilling the
Pauli's exclusion principle, their distribution functions could
slightly deviate from an exact Fermi-Dirac one.

Since we have provided a detailed account of non-extensive
statistics, and the reasons why we choose the analytical form of
the quantum distribution functions we adopted (together with its
derivation) in Paper I, we shall skip such topics here. We have
also considered in Paper I some physical reasons why to expect
that Boltzmann-Gibbs distribution functions could be considered as
an approximation. The same is valid for citations to previous
works, for which we adopted here the criterion to quote just those
needed for the explicit computations we are carrying on. This does
not mean that our references are the only works in cosmological
applications of non-extensivity, but only that for proper citation
of some of the others, we refer the reader to Paper I.

The layout of this work is as follows. Section 2 presents the
basics of the neutron to proton ratio in an evolving universe.
This section does not use much the fact that we are immersed in a
non-extensive setting, but just presents general results which are
valid disregarding the statistical mechanics used. Indeed, the
derivation being presented in Section 2 was already given by
others \cite{PHYSICA,Bernstein}, and we provide it here just for
the ease of discussion. In Sections 3 - 7 we give the details of
the analysis of the principle of detailed balance, and show how to
obtain a priori results on the possible range of physically
admitted values of $(q-1)$ without the need to compare with
experiments. Much of it is done in an analytical form, some is
solved numerically. In Section 8, we present a detailed comparison
between the two situations (full and approximate cases) that we
found possible for the principle of detailed balance. Towards the
end of this latter Section we provide a comparison with the latest
data available. In Section 9 we compute, for the first time in a
non-extensive framework, which is the modified capture time, the
time in which neutrons are captured into deuterons. Using this
result we are able to compute the primordial abundance of $^4$He
with a greater degree of precision than that obtained in all
previous works. We show that there are non-linear effects
introduced by the appearance of a slight non-extensivity. Finally,
we give some general discussion in our concluding remarks.

\section{The neutron to proton ratio}

We begin by turning again to the issue of the evolution of the
neutron abundance as the universe evolves. We shall base this
discussion in the work by, Bernstein, Brown and Feimberg
\cite{Bernstein}. As we have done before, we shall denote by
$\lambda_{pn}(T(t))$ the rate for the weak processes to convert
protons into neutrons and by $\lambda_{np}(T(t))$  the rate for
the reverse ones \cite{PHYSICA}. $X(T(t))$ will be, as usual, the
number of neutrons to the total number of baryons. For it, a valid
kinetic equation is
\begin{equation}
\label{eq:dX/dt}
\frac{dX(t)}{dt}=\lambda_{pn}(T)(1-X(t))-\lambda_{np}(T)X(t).
\end{equation} The solution to it is given by
\begin{equation}
\label{eq:XT} X(T)=\int_{t_0}^t dt' I(t,t') \lambda_{pn}(t') +
X(t_0)I(t,t_0).\end{equation} Here, $I(t,t')$ is
\begin{equation}
\label{eq:Itt'} I(t,t')=\exp \left( -\int_{t'}^t d\hat t
\Lambda(\hat t)\right),\end{equation} with
\begin{equation}
\Lambda(t)=\lambda_{pn}(t) + \lambda_{np}(t).\end{equation} Note
that this solution is completely general, and does not depend on
the statistical mechanics used, except by the implicit changes
introduced in the new reaction rates. As explained in
\cite{PHYSICA}, we simplify by taking $t_0=0$ and omitting the
$X(t_0)I(t,t_0)$ term. These approximations yield
\begin{equation}
\label{eq:Xt1} X(t)=\int_{0}^t dt' I(t,t')
\lambda_{pn}(t').\end{equation} Finally, we note that
\begin{equation}
I(t,t')=\frac{1}{\Lambda(t')}\frac{d}{dt'}I(t,t'),\end{equation}
or, equivalently,
\begin{equation}
\label{eq:Xt2} X(t)=\frac{\lambda_{pn}(t)}{\Lambda(t)} - \int_0^t
dt' I(t,t') \frac{d}{dt'} \left(
\frac{\lambda_{pn}(t')}{\Lambda(t')}\right).\end{equation}

To compute Eq. (\ref{eq:Xt2}), we need to know the reaction rates.
Let us consider $\lambda_{np}(t)$:
\begin{equation}
\label{eq:rnp} \lambda_{np}=\lambda_{\nu+n\rightarrow p+e^-} +
\lambda_{e^+n\rightarrow p+\bar \nu} + \lambda_{n\rightarrow
p+e^-+\bar \nu}\end{equation} that are individually given in
Ref.\cite{Weimberg}:
\begin{equation}
\label{eq:rnun,pe} \lambda_{\nu+n\rightarrow p+e^-}=A
\int_0^{\infty} dp_{\nu} p_{\nu}^2 p_e E_e
(1-f^e)f^{\nu},\end{equation}
\begin{equation}
\label{eq:ren,pnu} \lambda_{e^++n\rightarrow p+\bar \nu}=A
\int_0^{\infty} dp_e p_e^2 p_{\nu} E_{\nu}
(1-f^{\nu})f^e,\end{equation}
\begin{equation}
\label{eq:rn,penu} \lambda_{n\rightarrow p+e^-+\bar \nu}=A
\int_0^{p_0} dp_e p_e^2 p_{\nu} E_{\nu}
(1-f^{\nu})(1-f^e),\end{equation} with $A$ a constant, fixed by
the experimental value of  $\lambda_{n\rightarrow p+e^-+\bar
\nu}$, $p_{\nu,e}$ are the neutrino and electron momenta, and
$E_{\nu,e}$ their energies. In the energy domain we are
interested, some approximations are in order, see Refs.
\cite{PHYSICA,Bernstein} for discussion: 1) energy conservation is
$ E_{\nu}+m_n=E_e+m_p$, to be used in Eq. (\ref{eq:rnun,pe}) and
$E_{\nu}+m_p=E_e+m_n$, to be used in Eq.(\ref{eq:ren,pnu}). 2) In
Eq. (\ref{eq:rn,penu}), $E_{\nu}=\Delta m-E_e>0$, with $\Delta
m=m_n-m_p=1.29$MeV, from here comes the upper limit of the
integration range.  3) Pauli blocking factors are assumed equal to
1.

Since computations will involve the inverse processes of Eqs.
(\ref{eq:rnun,pe}) and (\ref{eq:ren,pnu}), we shall quote their
form below:
\begin{equation}
\label{eq:rep,nnu} \lambda_{e^-+p\rightarrow n+\nu}=A
\int_{p_e^0}^{\infty} dp_e p_e^2 p_\nu E_\nu
(1-f^{\nu})f^e,\end{equation}
\begin{equation}
\label{eq:rnup,ne} \lambda_{\bar \nu+p\rightarrow n+e^+}=A
\int_{\Delta m}^{\infty} dp_{\nu} p_{\nu}^2 p_e E_e
(1-f^e)f^{\nu},\end{equation} where $E_{\nu}+m_n=E_e+m_p$ for Eq.
(\ref{eq:rep,nnu}) and $E_{\nu}+m_p=E_e+m_n$ for Eq.
(\ref{eq:rnup,ne}). The lower limit of the integral
(\ref{eq:rep,nnu}) is given by the minimum momentum that electrons
must have in order to result $E_\nu>0$, $p_e^0=(\Delta
m^2-m_e^2)^{1/2}$. On the other hand, the lower limit in the
integral (\ref{eq:rnup,ne}) comes from the constraint
$E_\nu=E_e+\Delta m > \Delta m$, since we suppose that final
states are unbounded.

As remarked in \cite{PHYSICA}, in general, the electron and
neutrino temperatures, $T_e$ and $T_\nu$, may differ because at
the end of the freezing out period, electrons and positrons
annihilate, heating only the photons and maintaining with them
thermal equilibrium. Indeed, in Paper I we have computed the
amount of this deviation for our non-extensive setting. This
difference is, however, small, and we shall follow Bernstein et.
al. and set all temperatures equal, $T=T_e=T_\nu=T_\gamma$. In the
standard case this assumption ensure that the rates for reverse
reactions, such as $e^- + p \rightarrow n+ \nu$, obey the
principle of detailed balance (PDB).

\section{Principle of detailed balance}

The PDB states that if we know the expression for a reaction rate,
say $\lambda_{np}$, then this is related with the reaction rate
for the inverse reaction by an exponential factor,
\begin{equation}
\lambda_{pn}= e^{-\Delta m/T} \lambda_{np}.\end{equation} We shall
study what happen in the non-extensive case, by analyzing each
reaction appearing in Eq. (\ref{eq:rnp}) in a separate way.

\subsection{$\nu+n\leftrightarrow p+e^-$}

We would like to see what is the relationship between
$\lambda_{e^-+p\rightarrow n+\nu}^q$ and  $
\lambda_{\nu+n\rightarrow p+e^-}^q$, with
\begin{equation}
\lambda_{e^-+p\rightarrow n+\nu}^q=A\int_{p_e^0}^{\infty} dp_e
p_e^2 p_\nu E_\nu (1-f_q^\nu)f_q^e\end{equation} and
\begin{equation}
\lambda_{\nu+n\rightarrow p+e^+}^q=A\int_{0}^{\infty} dp_\nu
p_\nu^2 p_e E_e (1-f_q^e)f_q^\nu.\end{equation} Starting from
$\lambda_{e^-+p\rightarrow n+\nu}^q$, and taking into account that
$p_edp_e=E_edE_e$, $E_\nu=E_e-\Delta m$ and $E_\nu=p_\nu$, it is
possible to show that, using changes of variable,
\begin{equation}
\lambda_{e^-+p\rightarrow n+\nu}^q=A\int_{0}^{\infty} dp_\nu
p_\nu^2 p_e E_e (1-f_q^\nu)f_q^e.\end{equation} In the regime
$x=E/T \gg 1$, $f_q^i$ can be approximated by
\begin{equation}
\label{eq:fqi} f_q^i=e^{-x_i}+\frac{q-1}{2} x_i^2 e^{-x_i} \qquad
\textrm{with } x_i=E_i/T. \end{equation} Since during the period
of freezing, the temperature $T$ is low compared with the energies
appearing in the reaction rates, it is a good approximation to
neglect Pauli factors.  Then, to analyze the relationship between
\begin{equation}
\lambda_{e^-+p\rightarrow n+\nu}^q=A\int_{0}^{\infty} dp_\nu
p_\nu^2 p_e E_e f_q^e\end{equation} and
\begin{equation}
\label{eq:rnbfnun,pe} \lambda_{\nu+n\rightarrow
p+e^-}^q=A\int_{0}^{\infty} dp_\nu p_\nu^2 p_e E_e f_q^\nu,
\end{equation} it is enough to see how $f_q^e$ and $f_q^\nu$ relate
themselves. From Eq. (\ref{eq:fqi}), and recalling that in the
reactions $\nu+n\leftrightarrow p+e^-$, $E_\nu=E_e-\Delta m$, we
can directly write $f_q^e$ in terms of $f_q^\nu$. This yields
\begin{equation}
\label{eq:fefnurel} f_q^e= e^{-\Delta m/T}f_q^\nu + e^{-\Delta
m/T} \frac{q-1}{2} ( 2E_\nu \Delta m + \Delta m^2 )
\frac{e^{-E\nu/T}}{T^2}.\end{equation} Then, the PDB is no longer
valid, since the reaction rates in this case are related by
\begin{equation}
\label{eq:pdbq1} \lambda_{e^-+p\rightarrow n+\nu}^q= e^{-\Delta m
/T}\lambda_{\nu+n\rightarrow p+e^-}^q + \frac{q-1}{2} e^{-\Delta m
/T} \frac{A}{T^2} \int_0^{\infty} dp_\nu p_\nu^2 p_e E_e (2E_\nu
\Delta m+ \Delta m^2) e^{-E_\nu/T}\end{equation} with
$E_\nu=E_e-\Delta m$. We shall say that we are working within a
detailed balance with non-extensive corrections (BNE).

It is worth noting that we can obtain an approximate expression
among the rates, similar to what one obtains for the standard
situation. Indeed, using $E_\nu/T \gg 1$ in Eq.
(\ref{eq:fefnurel}), we get
\begin{equation}
\label{eq:pdbstf} f_q^e= e^{-\Delta m/T}f_q^\nu,\end{equation}
what immediately yields
\begin{equation}
\label{eq:pdbst1} \lambda_{e^-+p\rightarrow n+\nu}^q= e^{-\Delta m
/T}\lambda_{\nu+n\rightarrow p+e^-}^q. \end{equation} We can
recover the standard PDB, invoking this additional approximation,
what we shall refer as to standard PDB (BST). In what follows, we
shall show results both in the full BNE as in the BST
approximation, and we shall discuss the range of validity of Eq.
(\ref{eq:pdbstf}).

\subsection{$e^{+}+ n\leftrightarrow p+ \bar{\nu}$}

A similar analysis yields
\begin{equation}
\label{eq:pdbq2}\lambda_{e^++n\rightarrow p+\bar \nu}^q =
e^{-\Delta m /T}\lambda_{\bar \nu +p \rightarrow n+e^+}^q +
\frac{q-1}{2} e^{-\Delta m /T} \frac{A}{T^2} \int_0^{\infty} dp_e
p_e^2 p_\nu E_\nu (2E_\nu \Delta m+ \Delta m^2) e^{-E_e
/T},\end{equation} where $E_\nu=E_e+\Delta m$ and we neglected the
electron mass with respect to the energies. In fact, a careful
look at Eqs. (\ref{eq:pdbq1}) and (\ref{eq:pdbq2}) show that both
integrals are identical.

Within BST, the relationship between the rates again simplifies,
to give
\begin{equation}
\label{eq:pdbst2}\lambda_{e^++n\rightarrow p+\bar \nu}^q =
e^{-\Delta m /T}\lambda_{\bar \nu +p \rightarrow
n+e^+}^q.\end{equation}

\subsection{$n\rightarrow p+e^{-}+\bar{\nu}$}

Because of the fact that we have taken $1-f_q^i$ as $1$ in Eq.
(\ref{eq:rn,penu}), no modification arises in this case:
\begin{equation}
\label{eq:Ataurel} \frac{1}{\tau} = \lambda_{n \rightarrow p+e^- +
\bar \nu} = 0.0157 A \Delta m^5. \end{equation} This allow us to
determine $A$ in terms of a measurable quantity, the neutron mean
life $\tau$,
\begin{equation}
A= \frac{a}{\tau} \frac{1}{4} \Delta m^5,\;\;\;\;\;\;\;a=255.
\end{equation} We shall neglect the free decay of the neutron in this
section, particularly when computing $\lambda_{np}$. This makes
Eq. (\ref{eq:rnp}) to assume the form
$\lambda^q_{np}=\lambda^q_{\nu+n\rightarrow p+e^-} +
\lambda^q_{e^+n\rightarrow p+\bar{\nu}}$. Neglecting the blocking
factors in Eqs. (\ref{eq:rnun,pe}) and (\ref{eq:ren,pnu}) it is
evident that this two terms are identical. The same happens with
Eqs. (\ref{eq:rep,nnu}) and (\ref{eq:rnup,ne}), but in this case
we have to neglect, in addition, the electron mass in order to
have $p_0$ equal to $\Delta m$. Due to this,
$\lambda^q_{np}=2\lambda^q_{\nu+n\rightarrow p+e^-}$ and
$\lambda^q_{pn}=2\lambda^q_{e^-+p\rightarrow n+\nu}$.

Using all previous results, the valid relationship between
$\lambda^q_{np}$ and $\lambda^q_{pn}$, both in the BNE as in the
BST approximation are
\begin{equation}
\label{eq:pdbq} \lambda_{pn}^q= e^{-\Delta m /T}\lambda_{np}^q +
(q-1) e^{-\Delta m /T} I,\end{equation} where we have defined
\begin{equation}
\label{eq:I(nu)} I= \frac{A}{T^2}\int_0^{\infty} dp_\nu p_\nu^2
p_e E_e (2E_\nu \Delta m+ \Delta m^2) e^{-E_\nu/T}\end{equation}
with $E_e=E_\nu+\Delta m$, for BNE, and
\begin{equation}
\label{eq:pdbst} \lambda_{pn}^q= e^{-\Delta m /T}\lambda_{np}^q
\end{equation}
for the BST.

\section{Explicit form for the rates}

To compute  $\lambda^q_{np}$ and $\lambda^q_{pn}$, it is clear
that we have to know just one reaction rate, for instance
$\lambda^q_{\nu+n\rightarrow p+e^-}$, and the value of $I$ defined
in Eq. (\ref{eq:I(nu)}). We can solve for
$\lambda^q_{\nu+n\rightarrow p+e^-}$ using Eq.
(\ref{eq:rnbfnun,pe}). With the distribution functions being as in
Paper I we obtain
\begin{equation}
\lambda^q_{\nu+n\rightarrow p+e^-}=A\int_0^{\infty} dE_\nu E_\nu^2
(E_\nu+\Delta m)^2 f_q^\nu. \end{equation} These integrals are
easily computed changing variables to $x=E_\nu/T$ and using the
integral form of the function $\Gamma(n)$. Noting that
$\Gamma(n+1)=n!$ for $ n \in {\bf Z}$, defining $y=\Delta m/T$,
and using (\ref{eq:Ataurel}), we obtain
\begin{equation}
\label{eq:rnun,pe(y)} \lambda^q_{\nu+n\rightarrow p+e^-}=
\frac{1}{2} \frac{a}{\tau y^5} \left[12 + 6y + y^2 \right] +
\frac{a}{\tau y^5} (q-1)\left[180 + 60y + 6y^2\right].
\end{equation}

These same considerations are valuable to compute $I$. It is given
as
\begin{equation}
\label{eq:I(y)} I(y)= \frac{1}{2} \frac{a}{\tau y^5} \left[120y +
60y^2 + 12y^3 + y^4 \right]. \end{equation}

\subsection{$\lambda^q_{np}$}

We have already mentioned that $\lambda^q_{np}=
2\lambda^q_{\nu+n\rightarrow p+e^-}$, i.e.
\begin{equation}
\label{eq:rnp(y)} \lambda^q_{np}(y)=  \frac{a}{\tau y^5} \left[12
+ 6y + y^2 \right] + \frac{a}{\tau y^5} (q-1)\left[180 + 60y +
6y^2\right]. \end{equation} This equation is valid in both, the
full BNE and the BST approximation.
\begin{figure}
\vspace{-1.5cm}
\begin{center}
\includegraphics[width=8cm,height=11cm]{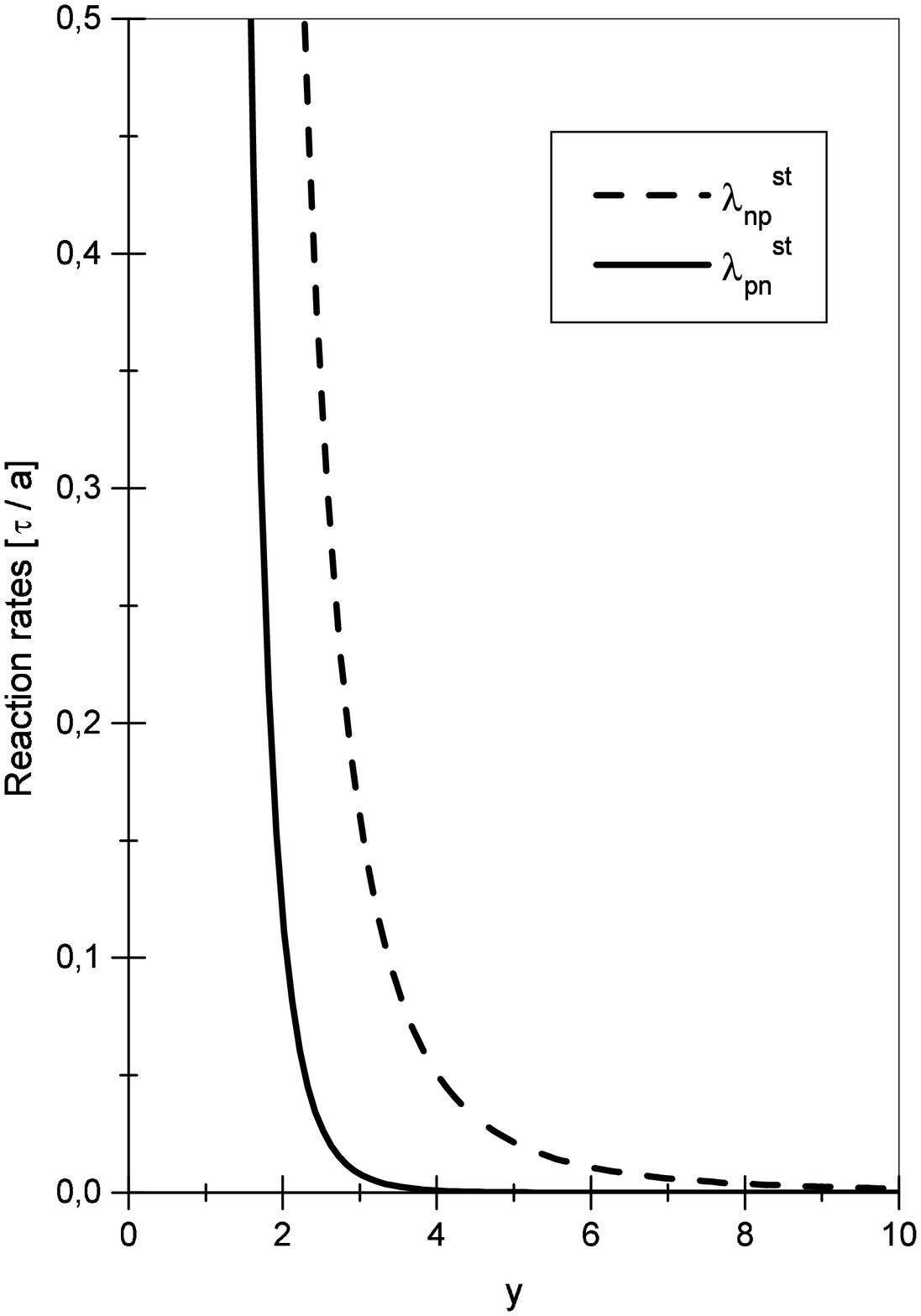}%
\includegraphics[width=8cm,height=11cm]{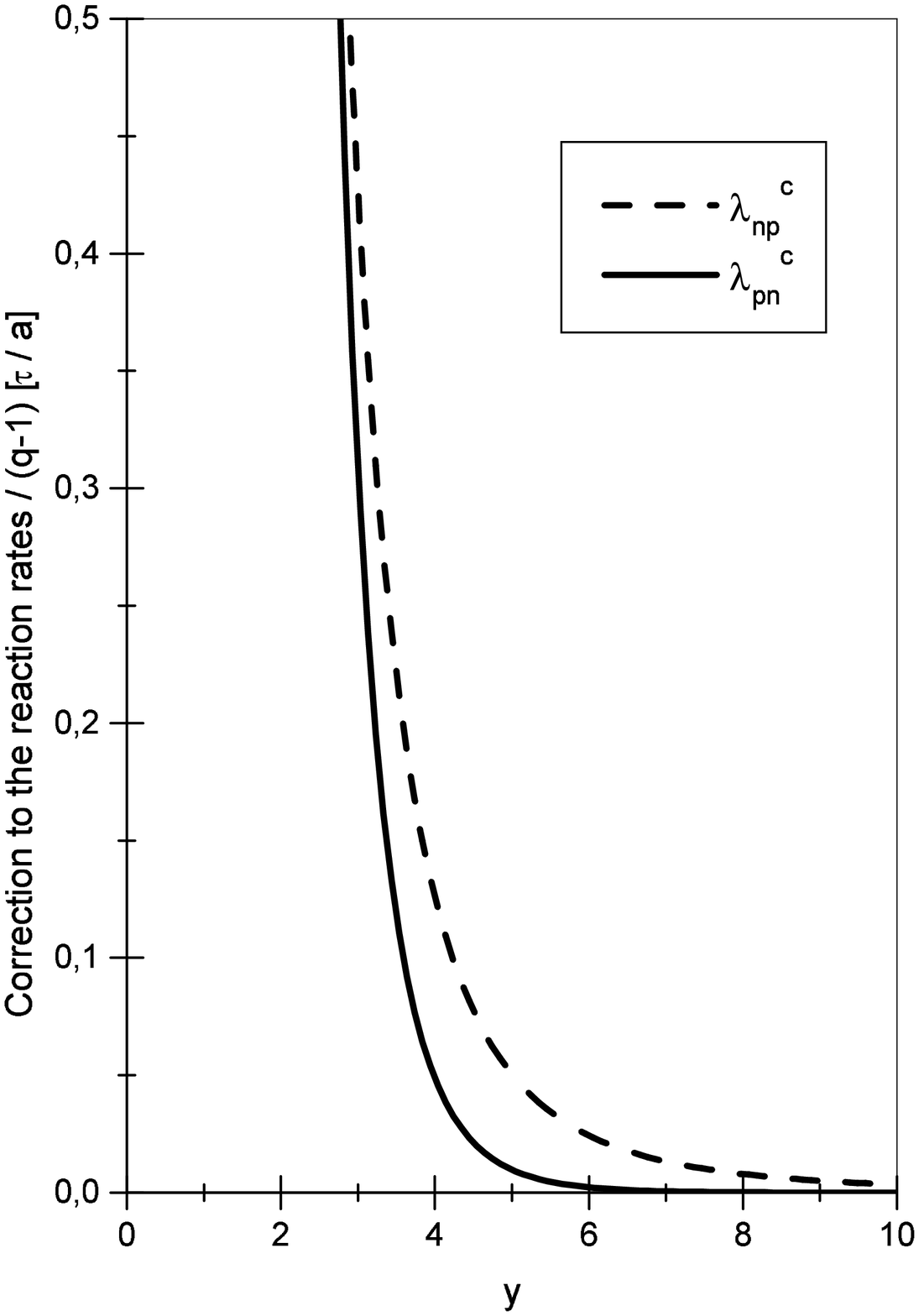}
\end{center}
\vspace{-1.5cm} \caption{Left: Standard reaction rates
$\lambda^{st}_{pn}(y)$ and $\lambda^{st}_{np}(y)$. Right:
Corrections produced by non-extensive statistics, normalized by
$(q-1)$, using the full BNE.} \label{fig:tasas}
\end{figure}

\subsection{$\lambda^q_{pn}$}

We use Eqs. (\ref{eq:pdbq}) and (\ref{eq:I(y)}) to get, within the
full BNE,
\begin{equation}
\label{eq:rpn(y)} \lambda^q_{pn}(y)= e^{-y} \lambda^q_{np}(y) +
\frac{q-1}{2} \frac{a}{\tau y^5} e^{-y} \left[120y + 60y^2 + 12y^3
+ y^4 \right]. \end{equation} It is also immediate to obtain this
rate in the BST, i.e.
\begin{equation}
\label{eq:rpnst(y)} \lambda^q_{pn}(y)= e^{-y}
\lambda^q_{np}(y).\end{equation} This last equation was used by
Torres and Vucetich \cite{PHYSICA}, where the BST was assumed from
the beginning.

It is worth noticing that since $\lambda^{st}_{pn}(y)$ and
$\lambda^{st}_{np}(y)$ have an asymptotic behavior analogous to
that of the standard rates, see Fig. \ref{fig:tasas}, also in this
case we expect that in the limit of low temperatures (i.e.
$y\rightarrow \infty$) the neutron abundance behaves as a
constant. That is, for low temperatures, the reaction rates are
practically zero and then, processes interchanging  neutrons into
protons and viceversa are negligible. In this sense,
$X^q(y\rightarrow \infty)$ stabilizes towards a constant.

\section{Evolution of the neutron abundance}

The formal solution for the evolution of the neutron abundance is
given by Eq. (\ref{eq:Xt2}). We shall use the variable $y=\Delta
m/T$, so that
\begin{equation}
\label{eq:Xqy1} X^q(y)= \frac {\lambda^q_{pn}(y)}{\Lambda^q(y)}-
\int_{0}^y dy^{\prime} I^q(y,y^\prime) \frac{d}{dy^{\prime}}
\left( \frac{\lambda^q_{pn}(y^{\prime})}{
\Lambda^q(y^{\prime})}\right), \end{equation} and the factor $I$
becomes
\begin{equation}
I^q(y,y^\prime) =\exp {\left( - \int_{y^\prime}^y    d\hat y
\left( \frac {d \hat t}{d \hat y} \right) \Lambda^q(\hat y)
\right)}. \end{equation}  To evaluate the Jacobian ${d \hat t}/{d
\hat y}$, we recall that the scale factor, $R$, in a
Friedmann-Robertson-Walker metric, goes as $R \simeq 1/T$,
independently of the statistics \cite{BARRACO}. Then, $\dot T /T =
- \dot R /R$, and the rhs is given by Einstein equations,
\begin{equation}
\frac {\dot R}{R} = \left( \frac{ 8 \pi G}{3} \rho_q
\right)^{1/2}. \end{equation} Here, $\rho_q$ is the energy density
of relativistic species, given in Paper I. When the universe is
dominated by  $e^-, e^+, \nu$ and $\gamma's$, and then $g_b=2$,
$g_f=2+2+2 \times 3=10$ and $g=\sum_b g^b+\frac{7}{8} \sum_f
g^f=43/4$, we obtain
\begin{equation}
\label{eq:eqr} \rho_q=\frac{\pi^2}{30} g T^4 + 35.85(q-1)T^4.
\end{equation}
We can now compute $dt/dy$,
\begin{equation}
\frac{dt}{dy}=\frac{dt}{dT} \frac{dT}{dy} = - \frac{1}{\dot T}
\frac{\Delta m}{y^2} . \end{equation} Since, $\dot T/T = - \dot R
/R$,
\begin{equation}
\dot T = - T \left( \frac{ 8 \pi G}{3} \rho_q \right)^{1/2},
\end{equation} and to first order in $(q-1)$,
\begin{equation}
\frac{1}{\dot T}= -\frac{1}{T^3}
\left(\frac{45}{4\pi^3Gg}\right)^{1/2}  \left[1 -\frac{15}{g\pi^2}
35.85(q-1) \right ]. \end{equation} In this sense, we see that
\begin{equation}
\label{eq:dt/dy} \frac{dt}{dy}=\frac{\tau}{a} b y \left[1 -c(q-1)
\right ],\end{equation} where $b$ and $c$ are given
\begin{equation}
b=\left(\frac{45}{4\pi^3Gg}\right)^{1/2} \frac{a}{\tau \Delta m^2}
,\;\;\;\;\;\;\;c=\frac{15}{g\pi^2} 35.85. \end{equation} Numerical
values for these constants can be obtained taking into account
that: $a=255$, $g=43/4$, $\Delta m=1.29$MeV, $\tau=887 \pm 2$s
\cite{Peacock} and $G=m_{pl}^{-2}$ and are $b=0.25$ and $c=5.07$.
This follows the treatment given in Ref. \cite{PHYSICA}.

Using all previous remarks,
\begin{equation}
\label{eq:Iyy'} I^q(y,y^\prime) =\exp {(K^q(y)-K^q(y^\prime))},
\end{equation} with $K^q(y)$ given by
\begin{equation}
K^q(y)= -\int d\hat y \left( \frac{d\hat t}{d\hat y}\right)
\Lambda^q(\hat y),\end{equation} or, equivalently, using Eq.
(\ref{eq:dt/dy}),
\begin{equation}
\label{eq:Kqy1} K^q(y)= -\frac{\tau b}{a} (1-c(q-1)) \int d\hat y
\hat y \Lambda^q(\hat y).\end{equation}

We then see that the needed steps to solve for the evolution of
$X^q$ as a function of $y$ are: 1) find the function $K^q(y)$ as
in Eq. (\ref{eq:Kqy1}), 2) substitute this result in the
expression (\ref{eq:Iyy'}) for $I^q(y,y')$, and 3) finally solve
the integral in (\ref{eq:Xqy1}). The complexity of all functions
involved requires a numerical procedure.  However, we can do
further before going to numerical work, since our aim is to
primarily see how the asymptotic value of $X$, $X^q(y\rightarrow
\infty)$, is affected as a function of the parameter $(q-1)$, and
ultimately, how this asymptotic value changes if one assume the
full BNE as compared with the approximate BST approach.

\section{Standard Balance: BST}
\label{sec:BST}

In this case,  $\lambda^q(y)/ \Lambda^q(y)$ coincides with its
standard analogous $\lambda^{st}(y)/ \Lambda^{st}(y)$. Indeed, by
definition, $\Lambda^q(y)=\lambda^q_{pn}(y)+\lambda^q_{np}(y)$,
and since BST establishes
$\lambda^q_{pn}(y)=e^{-y}\lambda^q_{np}(y)$, we obtain
$\lambda^q(y)/ \Lambda^q(y)=(1+e^{y})^{-1}=\lambda^{st}(y)/
\Lambda^{st}$.

We can get  $K^q(y)$ taking into account that, using Eqs.
(\ref{eq:rnp(y)}) and (\ref{eq:rpn(y)}),
\begin{equation}
\Lambda^q(y)=\frac{a}{\tau y^5}(1+e^y) \left[12 + 6y + y^2 \right]
+ (q-1)\left[180 + 60y + 6y^2\right]. \end{equation} Computing the
integral in Eq. (\ref{eq:Kqy1}),
\begin{equation}
\label{eq:Kqy2} K^q(y)= b(1-c(q-1))[K_1(y)+
\frac{q-1}{2}K_2(y)].\end{equation} with
\begin{equation}
\label{eq:K1} K_1(y)= \left[ \left(\frac{4}{y^3} + \frac{3}{y^2} +
\frac{1}{y} \right) + \left( \frac{4}{y^3} + \frac{1}{y^2} \right)
e^{-y} \right] \end{equation} and
\begin{equation}
\label{eq:K2} K_2(y)=\left[ \left(\frac{30}{y^3} + \frac{15}{y^2}
+ \frac{3}{y} \right) + \left( \frac{30}{y^3} +\frac{3}{y} \right)
e^{-y} - 3Ei(1,y) \right],\end{equation} and where the functions
$Ei(1,y)$ are
\begin{equation}
\label{eq:Ei} Ei(1,y)=\int_1^{\infty} \frac{e^{-yt}}{t}
dt.\end{equation} We recognize in Eq. (\ref{eq:Kqy2}) the standard
term $K^{st}(y)=bK_1(y)$. The presence of the factor $b(1-c(q-1))$
in the Jacobian $dt/dy$ produces the appearance in $K_2(y)$ of a
second order term in $(q-1)$. It is important to note that because
this function  diverges as $y^{-3}$ at the origin, it is not a
priori obvious that we can retain (as was done in \cite{PHYSICA}),
only the term linear in $(q-1)$. This is a key remark, and we
shall come back on this later.

We shall then try to obtain the whole solution, without
linearizing any of the functions involving $(q-1)$. Having shown
that $\lambda^q(y)/ \Lambda^q(y)=(1+e^{y})^{-1}$, we can write,
using Eq. (\ref{eq:Xqy1}), the asymptotic value $X^q(\infty)\equiv
\lim_{y \to \infty}X^q(y)$, as
\begin{equation}
\label{eq:Xqinf1} X^q(\infty)=\lim_{y \to \infty} e^{K^{q}(y)}
\int_{0}^{y} e^{-K^{q}(y')}
\frac{e^{y'}}{(1+e^{y'})^2}dy'.\end{equation} This expression
allows for an analytical study. Based on the divergent behavior of
$K^q$ at the origin and because of the good behavior of
\begin{equation}
\frac{d}{dy^{\prime}} \left( \frac{\lambda^{q}_{pn}(y^{\prime})}{
\Lambda^{q}(y^{\prime})}\right)=\frac{e^{y'}}{(1+e^{y'})^2},
\end{equation}
near zero, it is possible to discard \emph{a priori}, i.e. without
having the need to compare with any observation, the range of
$(q-1)$-values such that the non-extensive theory, considered
within the BST, looses its sense (i.e. the cases in which the
asymptotic values predicted for $X^q$ are out of the interval
[0,1/2]). Indeed, in order to avoid problems at the origin in Eq.
(\ref{eq:Xqinf1}), $K^q(y)$ has to be positive when $y\rightarrow
0$. Using Eqs. (\ref{eq:Kqy2}), (\ref{eq:K1}) and (\ref{eq:K2}),
this condition translates immediately into an allowed interval for
$(q-1)$, since
\begin{equation}
\lim_{y \to 0} K^q(y) \simeq \lim_{y \to 0} b(1-c(q-1)) \left[
\frac{4}{y^3} +\frac{4}{y^3} e^{-y} +
\frac{q-1}{2}\left(\frac{30}{y^3} +\frac{30}{y} e^{-y}\right)
\right],\end{equation} i.e.
\begin{equation}
\lim_{y \to 0} K^q(y) \simeq  b(1-c(q-1)) \left[ \frac{8}{y^3} +
(q-1)\frac{30}{y^3} \right].\end{equation} We see that, to have
$K^q(y)>0$, $(q-1)$ must be such that the inequality $-8/30 <
(q-1) < 1/c$ is fulfilled. Recalling that $c=5.07$ we see that
values that do not fulfill the condition
\begin{eqnarray}
\label{eq:rangoBST} -0.27 < (q-1) < 0.2
\end{eqnarray} must be automatically discarded. We mention in addition that
no matter the sign of $K^q$, the factor $e^{K(y)}$ multiplying the
integral in Eq. (\ref{eq:Xqinf1}) tends to 1 when  $y\rightarrow
\infty$.

\begin{figure}
\vspace{-1.5cm}
\begin{center}
\includegraphics[width=18cm,height=26cm]{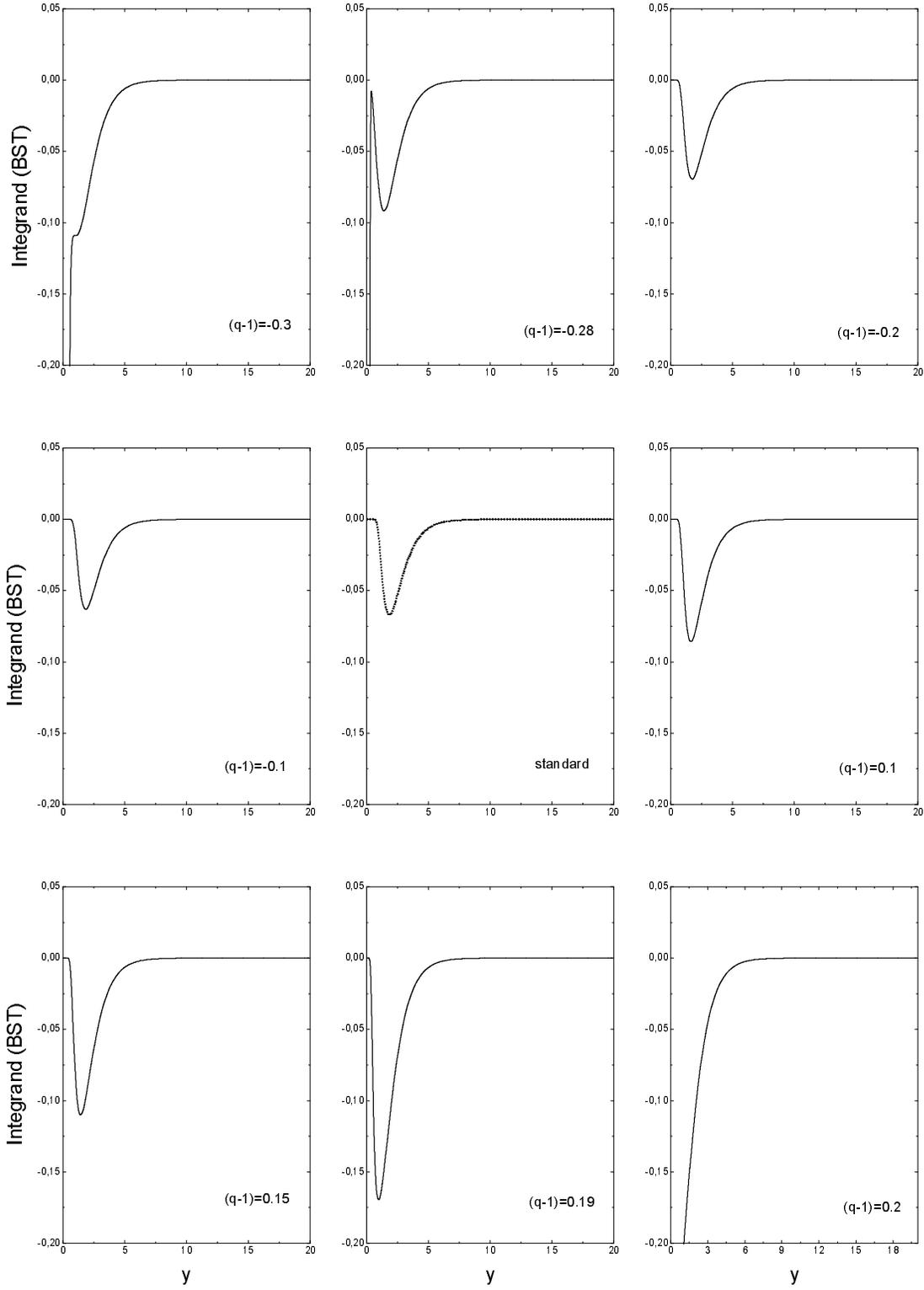}
\end{center}
\vspace{-3.5cm} \caption{ Behavior of the integrand needed in the
computation of $X^q(\infty)$ for different values of $(q-1)$
within the BST approach.} \label{fig:int-BST}
\end{figure}

\mbox{} Fig. \ref{fig:int-BST} shows the behavior of the integrand
of Eq. (\ref{eq:Xqinf1}) for different $(q-1)$-values. Note the
abrupt change of the plots when $(q-1)$ is near the extremes of
the range given by Eq. (\ref{eq:rangoBST}). The analytical
conclusion is then reinforced by these plots and is completely
confirmed when numerical computations are made. These latter show
that the integral (\ref{eq:Xqinf1}) grows without limit when
$(q-1)$ assumes values out of the range given by Eq.
(\ref{eq:rangoBST}).
We have then showed that there exist a range of  $(q-1)$ such that
the asymptotic value of $X^q(\infty)$ has physical sense. We can
now ask if there also exist a range of $(q-1) \in [-0.27,0.2]$
such that the $X^q(\infty)$ obtained to first order in $(q-1)$ is
consistent with the whole computation. This would be so if we can
prove that the first order result do not differ much from the real
result that we have already got.

\subsection{First order computation}

To compute  $X^q(\infty)$ to first order in $(q-1)$, we need to
write Eq. (\ref{eq:Xqinf1}) to first order and neglect quadratic
terms in Eq. (\ref{eq:Kqy2}). This gives
\begin{equation}
\label{eq:Kqy2ne} K^q(y)= K^{st}(y) +(q-1)K^c(y),\end{equation}
with
\begin{equation}
\label{eq:Kst} K^{st}(y)= b \left[ \left(\frac{4}{y^3} +
\frac{3}{y^2} + \frac{1}{y} \right) + \left( \frac{4}{y^3} +
\frac{1}{y^2} \right) e^{-y} \right],\end{equation} and
\begin{equation}
\label{eq:Kc} K^c(y)=\frac{b}{2}\left[ \left(\frac{30}{y^3} +
\frac{15}{y^2} + \frac{3}{y} \right) + \left( \frac{30}{y^3}
+\frac{3}{y} \right) e^{-y} - 3Ei(1,y)
\right]-cK^{st}.\end{equation} We write $e^{K^{q}}$ to first order
as
\begin{equation}
e^{K^{q}}=e^{K^{st}} e^{(q-1)K^c}\simeq e^{K^{st}}
[1+(q-1)K^c].\end{equation} Substituting this expansion into
(\ref{eq:Xqinf1}) and retaining only the linear term we get
\begin{equation}
\label{eq:XqinfO(q-1)} X^q(\infty)=  X^{st}(\infty)+ (q-1)
X^c(\infty),\end{equation} with
\begin{equation}
X^{st}(\infty)=\lim_{y \to \infty} e^{K^{st}(y)} \int_0^{y}
e^{-K^{st}(y')} \frac{e^{y'}}{(1+e^{y'})^2}dy' ,\end{equation} and
\begin{equation}
X^c(\infty)=\lim_{y \to \infty} e^{K^{st}(y)} \int_0^{y}
e^{-K_{st}(y')} [K^c(y)-K^{c}(y')] \frac{e^{y'}}{(1+e^{y'})^2} dy'
.
\end{equation} This integrals can be computed numerically, and the result is
\begin{equation}
\label{eq:XqO(q-1)} X^q(\infty)=0.15+(q-1)0.18\end{equation} where
the value $ X^{st}(\infty)=0.15$ is the standard one.
\begin{figure}
\vspace{-2cm}
\begin{center}
\includegraphics[width=11cm,height=12cm]{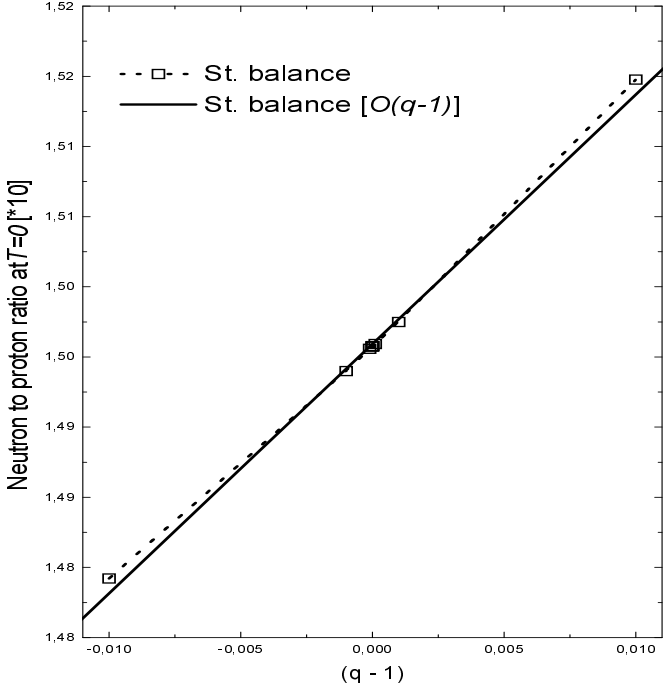}
\end{center}
\vspace{-1.5cm} \caption{ Comparison between $X^q(\infty)$ as a
function of $(q-1)$,  to first order, and using the full numerical
solution, both obtained within the BST approach.}
   \label{fig:fit-pdbt}
\end{figure}
\mbox{}Fig. \ref{fig:fit-pdbt} shows the linear dependence of Eq.
(\ref{eq:XqO(q-1)}), together with the full solution given by Eq.
(\ref{eq:Xqinf1}). Although within the framework of the asymptotic
approach to quantum distribution functions, an analogous equation
to (\ref{eq:XqO(q-1)}) was presented already by some of us
\cite{PHYSICA}, we have now shown, for the first time, what is
behind that approach, what sustains it, and the range of its
validity.

\section{Non-extensive balance: BNE }
\label{sec:BNE}

In this section we shall solve Eq. (\ref{eq:Xqy1}) without
approximations. That is, we shall take for $\lambda^q_{pn}$ and
$\lambda^q_{np}$, the expressions given by (\ref{eq:rnp(y)}) and
(\ref{eq:rpn(y)}), respectively. Our aim will be to quantify the
amount of the deviation that it is produced when the BST, instead
of the full BNE, is considered.

By definition, $\Lambda^q(y)=\lambda_{pn}^q(y) +
\lambda_{np}^q(y)$, adding  (\ref{eq:rnp(y)}) and
(\ref{eq:rpn(y)}) we see that $\Lambda^q(y)$ is given by the sum
of two terms
\begin{equation}
\Lambda^{st}(y)=\frac{a}{\tau y^5} (12+6y+y^2)
\left[1+e^{-y}\right],\end{equation}
\begin{equation}
\Lambda^c(y)=\frac{1}{2} \frac{a}{\tau y^5} \left[ 90 \left[1 +
e^{-y}\right] + 30 y \left[ 1 + 5 e^{-y}\right] + y^2 \left[1 + 21
e^{-y}\right] + 12 y^3 + y^4 \right].\end{equation}

\begin{figure}
\vspace{-1.5cm}
\begin{center}
\includegraphics[width=18cm,height=26cm]{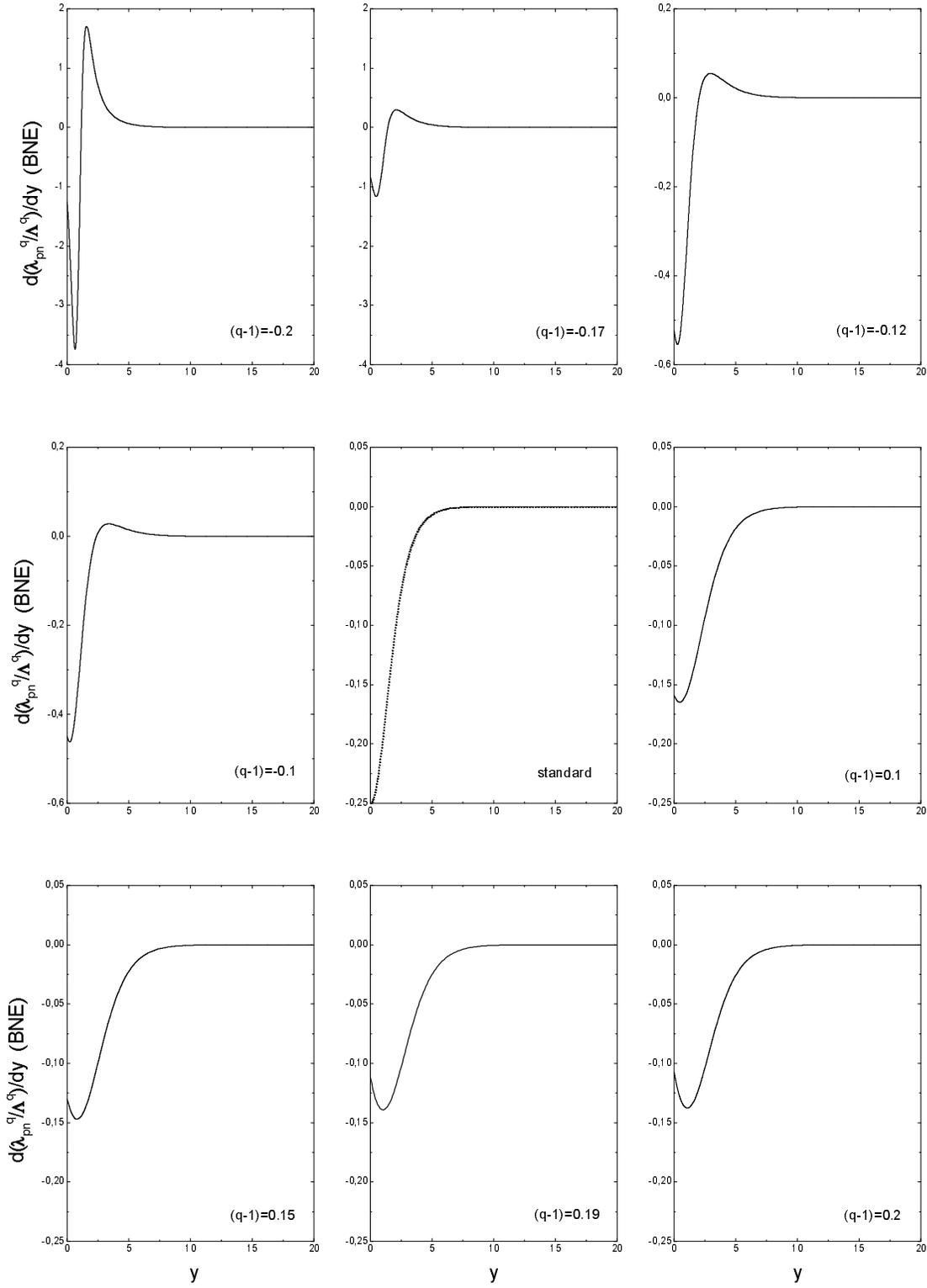}
\end{center}
\vspace{-3.5cm} \caption{ Behavior of the derivative of
$\lambda^q_{pn}(y)/\Lambda^q(y)$ for different $(q-1)$-values
within the BNE.} \label{fig:dab-real}
\end{figure}

\begin{figure}
\begin{center}
\vspace{-1.5cm}
\includegraphics[width=18cm,height=26cm]{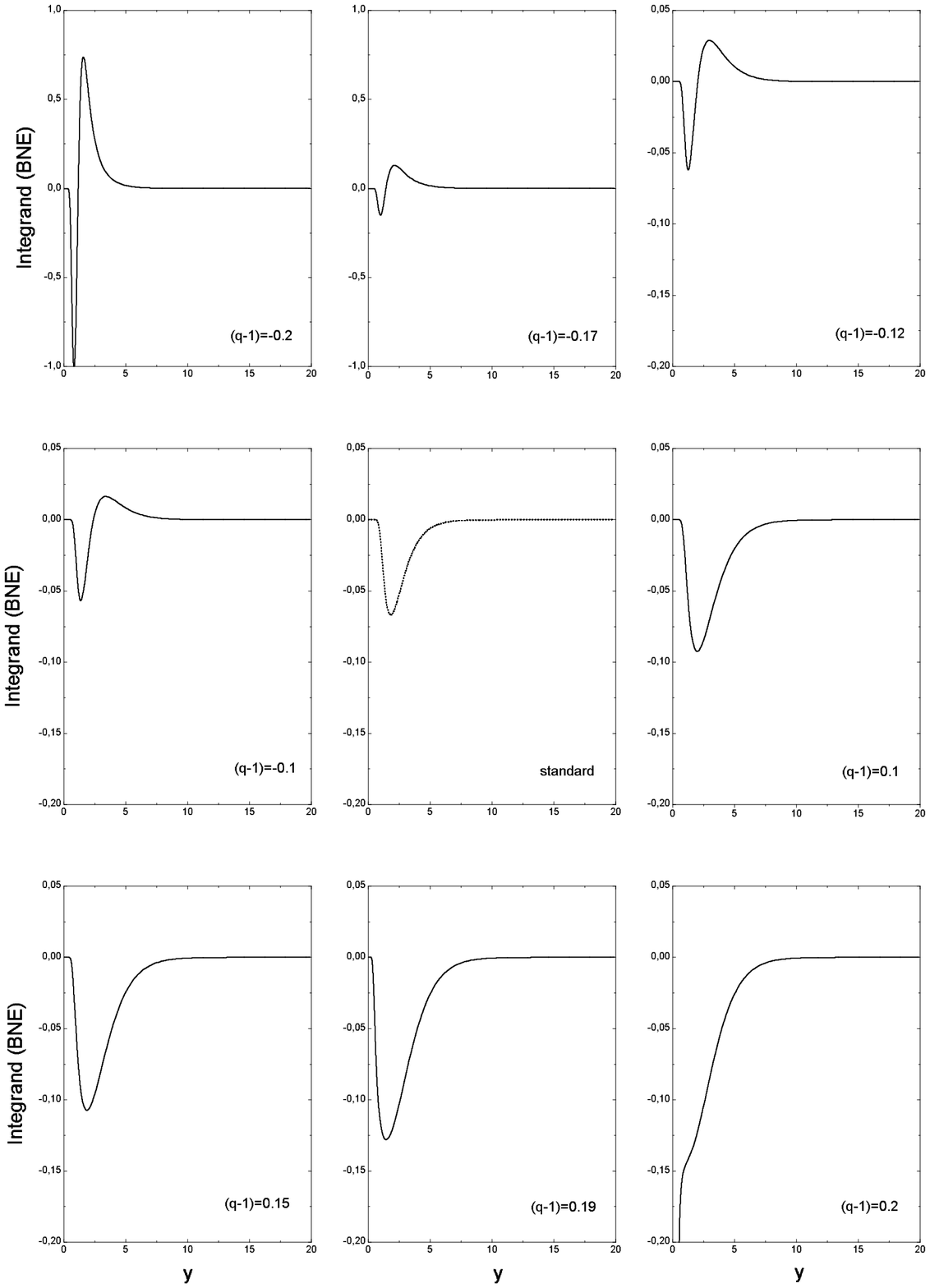}
\end{center}
\vspace{-3.5cm} \caption{Behavior of the integrand involved in the
computation of $X^q(\infty)$ for different $(q-1)$-values, within
the BNE.} \label{fig:int-real}
\end{figure}

Because of the linearity of the integral operator, the same
happens to $K^q(y)$ in Eq. (\ref{eq:Kqy1}). Using the previous
expressions, we get
\begin{equation}
\label{eq:hKqy2} K^q(y)= b(1-c(q-1))[\hat K_1(y)+ \frac{q-1}{2}
\hat K_2(y)].\end{equation} with
\begin{equation}
\hat K_1(y)= \left[ \left(\frac{4}{y^3} + \frac{3}{y^2} +
\frac{1}{y} \right) + \left( \frac{4}{y^3} + \frac{1}{y^2} \right)
e^{-y} \right] \end{equation} and
\begin{equation}
\hat K_2(y)=\left[ \left(\frac{30}{y^3} + \frac{15}{y^2} +
\frac{3}{y} \right) + \left( \frac{30}{y^3} + \frac{60}{y^2} +
\frac{3}{y} \right) e^{-y} -y -12\ln y - 3Ei(1,y)
\right]\end{equation} where $Ei(1,y)$ were given in (\ref{eq:Ei}).
As before, we do recognize the standard term $K^{st}(y)=b\hat
K_1(y)$. Also here, within the BNE, it happens that
\begin{equation}
\lim_{y \to \infty}
\frac{\lambda^q_{pn}(y)}{\Lambda^q(y)}=0.\end{equation} Indeed,
since within the BNE, in the limit  $y\rightarrow \infty$ we have
$\lambda^q_{pn}(y) \propto y^{-3}$ and $\lambda^q_{np}(y) \propto
y^{-1}e^{-y}$, we obtain
\begin{equation}
\lim_{y \to \infty} \frac{\lambda^q_{pn}(y)}{\Lambda^q(y)}=\lim_{y
\to \infty}\frac{1}{1+\lambda^q_{pn}(y)/\Lambda^q(y)} = \lim_{y
\to \infty} \frac{1}{1+e^y}=0.\end{equation} We then obtain an
analogous to Eq. (\ref{eq:Xqinf1})
\begin{equation}
\label{eq:XqinfBNE} X^q(\infty)= - \lim_{y \to \infty} e^{K^q(y)}
\int_0^{y} e^{-K^q(y')} \frac{d}{dy'}
\left(\frac{\lambda^q_{np}(y)}{\lambda^q_{pn}(y)}\right)
dy'.\end{equation} It is important to note that $K^q(y)$ diverges
as $y^{-3}$ at the origin, and as $-y$ at infinity. Then, first
order developments will not do. The complex dependencies of the
integrand of Eq. (\ref{eq:Xqy1}) with $(q-1)$ makes much harder to
a priori analyze the validity range, as done within the BST. We
can, however, make a detailed analysis of the behavior of the
derivative of the function $\lambda^q_{pn}(y)/\Lambda^q(y)$ and of
the integrand of Eq. (\ref{eq:Xqy1}), for different
$(q-1)$-values. We show this in Figs. \ref{fig:dab-real} and
\ref{fig:int-real}.
\mbox{} From the analysis of these figures we see that the BST and
the BNE differ in a fundamental way, which translates into the
value of  $X^q(\infty)$. Within the BNE, and because of the
behavior of the derivative of the function
$\lambda^q_{pn}(y)/\Lambda^q(y)$ for negative values of $(q-1)$,
it is not guaranteed that the integrand in Eq. (\ref{eq:Xqy1}) is
negative. For $(q-1)<-0.1$ the integral of (\ref{eq:Xqy1}) is
positive (i.e. $X^q(\infty)<0$), and this discards values of $q$
such that $(q-1)<-0.1$. On the other hand, looking at Fig.
\ref{fig:dab-real} for positive values of $(q-1)$, we see that the
upper bound is between 0.19 and 0.2, in this case similarly to the
BST. Then, for the BNE, the values for which the theory has
physical sense a priori of any experimental comparison are
\begin{equation}
\label{eq:rangoBNE} -0.1< (q-1) < 0.2.
\end{equation} The BNE is more restrictive than the BST. Out of this range,
the physical sense of the description is lost.

\section{Comparing both approaches, and bounding the value of $q$}

\begin{figure}
\vspace{-2cm}
\begin{center}
\includegraphics[width=11cm,height=12cm]{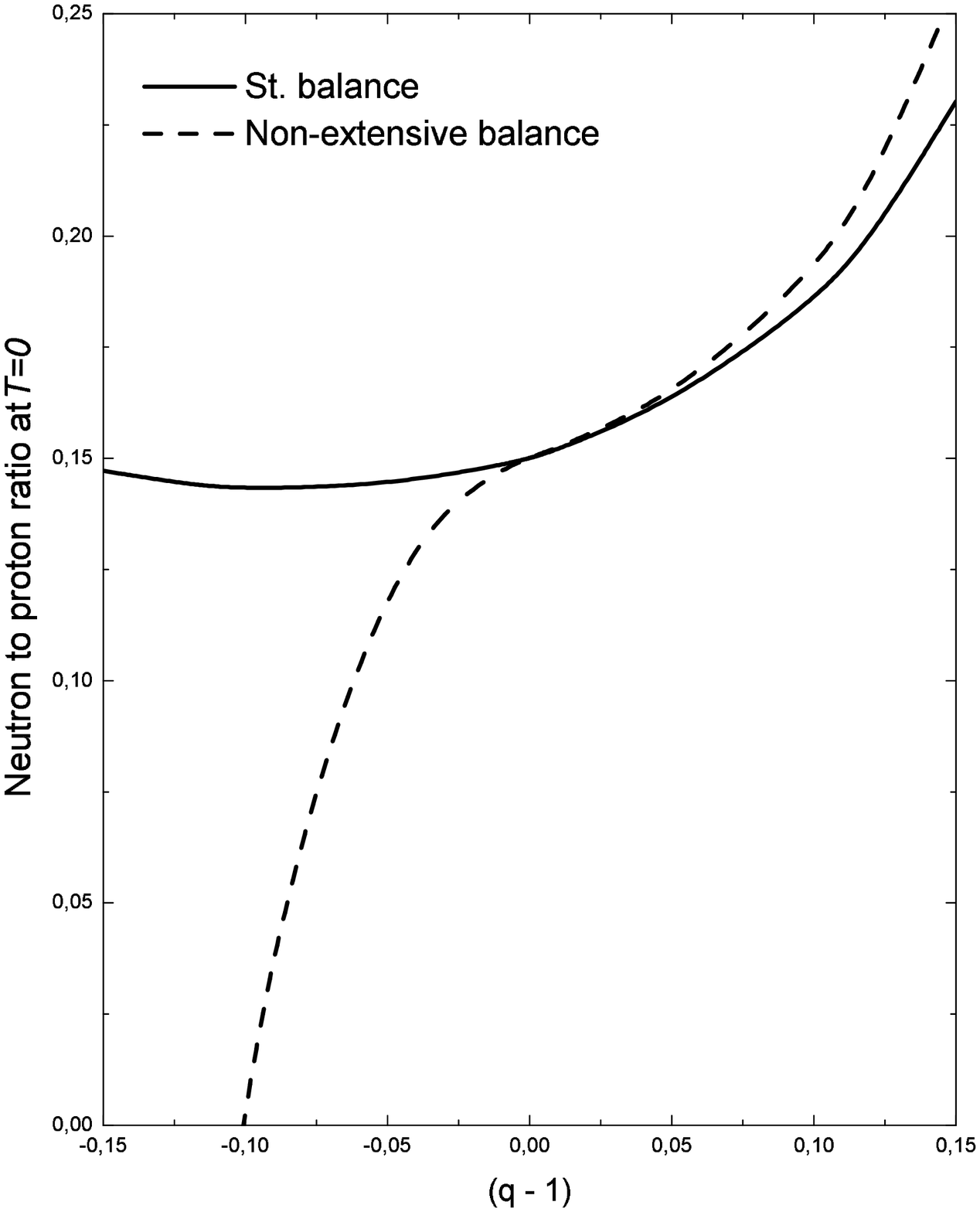}
\end{center}
\vspace{-1.5cm} \caption{$X^q(\infty)$ as a function of $(q-1)$,
both for the full BNE and the BST.} \label{fig:comp}
\end{figure}

We now compare the results obtained in Sections \ref{sec:BST} and
\ref{sec:BNE}. Fig. \ref{fig:comp} shows $X^q(\infty)$ for
different values of $(q-1)$, both in the full BNE as well as in
the BST approximation. Differences are notable, and they grow with
increasing values of  $|q-1|$, as it tends to its limiting values.
Table 1 gives the explicit values of $X^q(\infty)$ obtained using
a numerical integration scheme for the BST and for the full BNE,
and the analytical results of the first order approximation for
the BST. For values of $(q-1)$ near 0, the BST numerical
computation is perfectly described by the linear approximation.
\begin{table}
\caption{Comparison of $X^q(\infty)$-values as a function of
$(q-1)$ for the different approaches.}
\begin{center}
\begin{tabular}{|r|c|c|c|}
\hline
 $(q-1)$ & $X^q$ (BST) & $X^q$ (BST) & $X^q$ (BNE)\\
         &             & ($O$ $(q-1)$)&   \\
\hline \hline
 -0.150   & 0.1472 & 0.1230 & -0.6571 \\
 -0.120   & 0.1441 & 0.1284 & -0.1061 \\
 -0.100   & 0.1431 & 0.1320 &  0.0261 \\
 -0.050   & 0.1440 & 0.1410 &  0.1299 \\
 -0.010   & 0.1484 & 0.1482 &  0.1476 \\
 -0.001   & 0.1499 & 0.1498 &  0.1498 \\
  0.0\hspace{0.4cm}  & 0.1500 & 0.1500 &  0.1500 \\
  0.001   & 0.1502 & 0.1502 &  0.1503 \\
  0.010   & 0.1520 & 0.1518 &  0.1523 \\
  0.050   & 0.1625 & 0.1590 &  0.1635 \\
  0.100   & 0.1851 & 0.1680 &  0.1915 \\
  0.120   & 0.1991 & 0.1716 &  0.2109 \\
  0.150   & 0.2303 & 0.1770 &  0.2561 \\
\hline
\end{tabular}
\end{center}
\label{tab:tabla1}
\end{table}
The deviation from the standard case is greater for the BNE than
for the BST, for equal values of $(q-1)$, i.e. the BST
underestimates the effect of non-extensivity. This can be seen
directly from Fig. \ref{fig:comp}, or equivalently from Table 2,
where we show the coefficients of the linear fit of $X^q(\infty)$
near 0, both for the full BNE and the BST. In the sake of
completeness, we also show the first order result for the BST
(this is not a fit, but comes directly from the analytical
computations shown before).
\begin{table}
\caption{Coefficients of the linear  fits of $X^q(\infty)$ near
$(q-1)=0$ for the full BNE, the BST approximation, and for the
first order computation of the BST.}
\begin{center}
\begin{tabular}{|l|c|c|}
\hline
 Case & $X^{st}$ & $X^c$ \\
\hline \hline
 BST Approximation & 0.15 & 0.18\\
 BST $O(q-1)$ & 0.15 & 0.18\\
 BNE Full & 0.15 & 0.23\\
\hline
\end{tabular}
\end{center}
 \label{tab:tabla2}
\end{table}

Until now, we have discussed the physical viability of the
statistical description, establishing a range of a priori
discarded values of $(q-1)$ both for the full BNE case, the real
situation, and for the BST approximation. These values are those
for which $X^q(\infty)$ is $<0$ or $>0.5$. This can be done before
than the comparison with any given observational data or
experiment. Recall that $X(\infty)$ is the neutron to proton
ratio, for which clearly, a negative value has no sense. In
addition, $X(\infty)>0.5$ is in conflict with the reaction rates
dependence with temperature, since  $X=0.5$ is the initial
condition for the kinetic equation (\ref{eq:dX/dt}) when
$T\rightarrow \infty$ (i.e. $y\rightarrow 0$). Within the range of
physical viability we now look for consistency with observations.
This will further restrict the range of admitted values of
$(q-1)$. To obtain a direct bound upon $(q-1)$ using the
primordial abundance of $^4$He it is necessary to study in detail
the free decay of neutrons, happening  between the moment of
freezing out of the weak interactions ($t\simeq$ 1 s) and the
moment in which the temperature of the universe is similar to the
binding energy of D ($t\simeq$ 3 minutes). We shall provide, for
the first time in a non-extensive setting, a detailed account of
this in the following section, but nevertheless, let us give here
some preliminary considerations in the sense of Ref. \cite{PRL}.

Including the effects of the neutron decay in the equation for the
evolution of $X^q(t)$ we have,
\begin{equation}
\label{eq:Xdecay} X^q(t)=\exp(-t/\tau) \bar{X}^q(t),
\end{equation} where $\bar X^q$ is the already obtained ratio and
$\tau$ is the neutron mean life. In the capture time, $t=t_c$,
when the temperature is below the D binding energy (2.23 MeV),
neutrons are captured into D. Then, almost all neutrons present at
$t=t_c$ are converted into $^4$He. Substituting the value of $t_c$
into Eq. (\ref{eq:Xdecay}) and using the value of
$\bar{X}^q(\infty)$ previously found, we shall obtain half of the
mass fraction of all $^4$He primordially produced. For the time
being, we shall adopt the standard value for $\exp(t_c/\tau)$,
which is $ \simeq 0.8$ \cite{Bernstein}. With this we obtain,
\begin{eqnarray} \nonumber Y^q_p &\equiv& 2X_4^q\simeq e^{-t_c/\tau}
2\bar{ X}^q(\infty) \\ \nonumber & = & 0.8\times 2 \times [0.15+(q-1)0.23]\\
\label{eq:Yqp}& = &  0.24 + (q-1)0.37.
\end{eqnarray}
There is no absolute consensus about the observational value of
$^4$He. The two greatest compilations give  \cite{Beuler},
\begin{equation}
Y_p=0.244\pm 0.004 \qquad \textrm{and} \qquad Y_p=0.234\pm 0.004.
\end{equation}
To consider a typical case, we average over the two mean values
and twice the error bar,
\begin{equation}
Y_p=0.239\pm 0.008.
\end{equation}
If at the same time we neglect the difference between the standard
theoretical and the observational values, which is given by 0.001,
in order to obtain a first bound,  we can get, using Eq.
(\ref{eq:Yqp}) and asking for $|q-1|0.37<0.008$,
\begin{equation}
|q-1|<0.021.\end{equation} Within the most complete treatment of
the principle of detailed balance (BNE), which accounts for the
more in depth analytical study on the influence of a slight
non-extensivity in primordial nucleosynthesis, the value of $q$
could differ from 1 at the level of a few percent and still be in
agreement with current constraints. See below for a more detailed
treatment.

\section{The capture time}

The aim of this section is to show how the capture time $t_c$ is
modified in the non-extensive framework. Recalling that at early
times $\dot R/R=(8 \pi \rho_q/3 m_{pl})^2$ and
$R(t)T_\nu(t)=constant$ after the $e^+e^-$ annihilation, it is
straightforward to write
\begin{equation}
t= \left(\frac{3}{8\pi G}\right)^{1/2} \int_{T_\nu}^\infty
\rho_q^{-1/2} \frac{dT'_\nu}{T'_\nu} + t_0,
\end{equation}
where $t_0$ is a constant whose standard value is of the order 2
seconds, see \cite{Bernstein}. In Paper I, it was shown that the
energy density could be written as $ \rho_R = (\pi^2/30) g_*^q
T^4$. Taking also into account that the photon and neutrino
temperatures are related by $T_\gamma=(11/4)^{1/3}[1+0.109(q-1)]
T_\nu$, we can solve the definite integral to obtain:
\begin{equation}
\label{eq:tTrel} t=\left( \frac{45}{16\pi^3 g_{eff}}\right)^{1/2}
\left(\frac{11}{4}\right)^{2/3}\frac
{m_{pl}}{T^2_\gamma}[1-5.06(q-1)] + t_0,
\end{equation}
where $g_{eff}=2(11/4)^{4/3}+21/4 \simeq 12.95$.

In equilibrium, neutrons, protons and deuterons behave as free
non-relativistic gases, with number densities given by
\begin{equation}
\label{nr} n_q^i=g_i \left(\frac{m_i T_\gamma}{2\pi}\right)^{3/2}
e^{-(m_i-\mu_i)/T_\gamma} \times \left[ 1 + \frac{q-1}{2} \left(
\frac {15}4 + 3 \frac {m_i-\mu_i}{T_\gamma} + \left(\frac
{m_i-\mu_i}{T_\gamma}\right)^2 \right) \right].
\end{equation}
Here, the subscript $i$ stands for $n,p$ and $D$, for neutrons,
protons and deuterons respectively, $g_n=g_p=2, g_D=3$ and
$\mu_i,m_i$ are the chemical potentials and masses of the
particles. At early times, these gases are in chemical
equilibrium, so that $\mu_D=\mu_n+\mu_p$. Therefore,
\begin{equation}
\label{eq:SahaD} \frac{n_q^n n_q^p}{n_q^D}=\frac{g_n g_p}{g_D}
\left(\frac{m_n m_p}{m_D} \right)^{3/2}
\left(\frac{T_\gamma}{2\pi} \right)^{3/2} e^{-\epsilon_D/T_\gamma}
\left[\frac{\bar u((m_n-\mu_n)/T)\bar u((m_p-\mu_p)/T)}{\bar
u((m_D-\mu_D)/T)}\right],
\end{equation}
with $\epsilon_D=m_n+m_p-m_D$, and where we have made use of the
definition,
\begin{equation}
\label{ubar} \bar u((m_i-\mu_i)/T)\equiv 1 + \frac{q-1}{2} \left(
\frac {15}4 + 3 \frac {m_i-\mu_i}{T} + \left(\frac
{m_i-\mu_i}{T}\right)^2 \right).
\end{equation}
Equation (\ref{eq:SahaD}) is the generalized Saha equation that
describes the deuterium formation (see Paper I for a general
derivation). Helium appearance is inhibited by the ``Deuteron
bottleneck'', represented by the reactions:
\begin{eqnarray}
\label{eq:reacD1}
n+p & \rightarrow & \rm{D}+\gamma,\\
\label{eq:reacD2}
\rm{D}+\rm{D}& \rightarrow & \rm{T}+p, \\
\label{eq:reacD3} \rm{D}+\rm{T}& \rightarrow & ^4\!\rm{He} +n.
\end{eqnarray}
Eq. (\ref{eq:SahaD}) can be given in terms of the normalized
abundance fractions $X_q^i=n_q^i/n_q^B$. To this purpose we
introduce the baryon to photon ratio $\eta_q=n_q^B/n_q^\gamma$ and
recall that
$n_q^\gamma=[2\zeta(3)+3!\zeta(4)(q-1)](T^3_\gamma/\pi^2)$. We get
for the quantity $G^q_{np}=X_q^n X_q^p/ X_q^D$, to first order in
$(q-1)$, the following result
\begin{equation}
\label{eq:Gqnp} G^q_{np}=\frac{\pi^{1/2}}{\zeta(3)}
\frac{2}{3\eta_q} \left(\frac{m_pm_n}{2m_DT\gamma}\right)^{3/2}
e^{-\epsilon_D/T_\gamma} \left[1+\frac{q-1}{2}(u_n+u_p-u_D)
\right].
\end{equation}
Here, as in Paper I, $u_i=15/4+3(m_i-\mu_i)/T+((m_i-\mu_i)/T)^2$.
As in the standard case analyzed in Ref. \cite{Bernstein}, to
determine the neutron capture time $t_c$ we need to examine the
sequence of reactions (\ref{eq:reacD1}-\ref{eq:reacD3}). It is
convenient to do so in terms of an scaled temperature
$z=\epsilon_D/T_\gamma$, so that they involve rate parameters of
the form
\begin{equation}
R^q=\frac{dt}{dz}\langle \sigma v \rangle_T n_q^B,
\end{equation}
where $\langle \sigma v \rangle_T$ denotes the thermal average of
the relevant cross section times the relative velocity
\cite{Bernstein}. Using Eq. (\ref{eq:tTrel}) and writing the
baryon number density in terms of the baryon to photon number
density ratio, $\eta_q$, we get
\begin{equation}
\label{eq:Rq} R^q=\frac{\eta_q}{z^2} \left(\frac{45}{\pi^7
g_{eff}}\right)^{1/2} \left(
\frac{11}{4}\right)^{2/3}\zeta(3)\epsilon_D m_{pl} \langle \sigma
v \rangle_T [1+0.34(q-1)].
\end{equation}
Taking $\eta_0=3.57 \times 10^{10}$ \cite{Peacock} as a nominal
value for $\eta_q$ and considering $\langle \sigma_{np} v
\rangle_T=4.6 \times 10^{-20}$ \cite{Peebles} we obtain
\begin{equation}
\label{eq:Rqnp} R^q_{np}=3.63 \left(\frac{29}{z^2}\right)^{2}
\frac{\eta_q}{\eta_0}[1+0.34(q-1)].
\end{equation}

\begin{figure}[t]
\vspace{-2cm}
\begin{center}
\includegraphics[width=8cm,height=11cm]{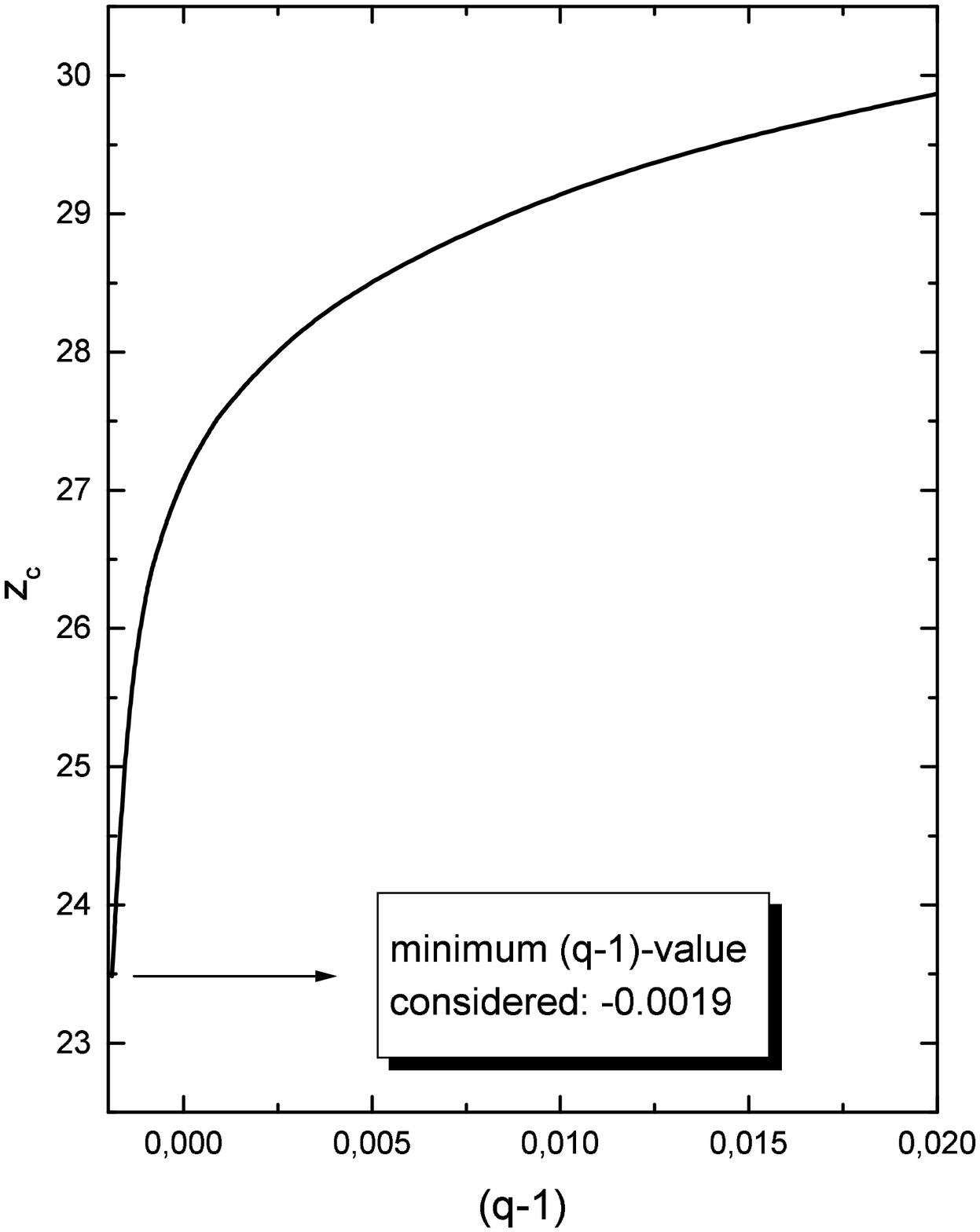}
\includegraphics[width=8cm,height=11cm]{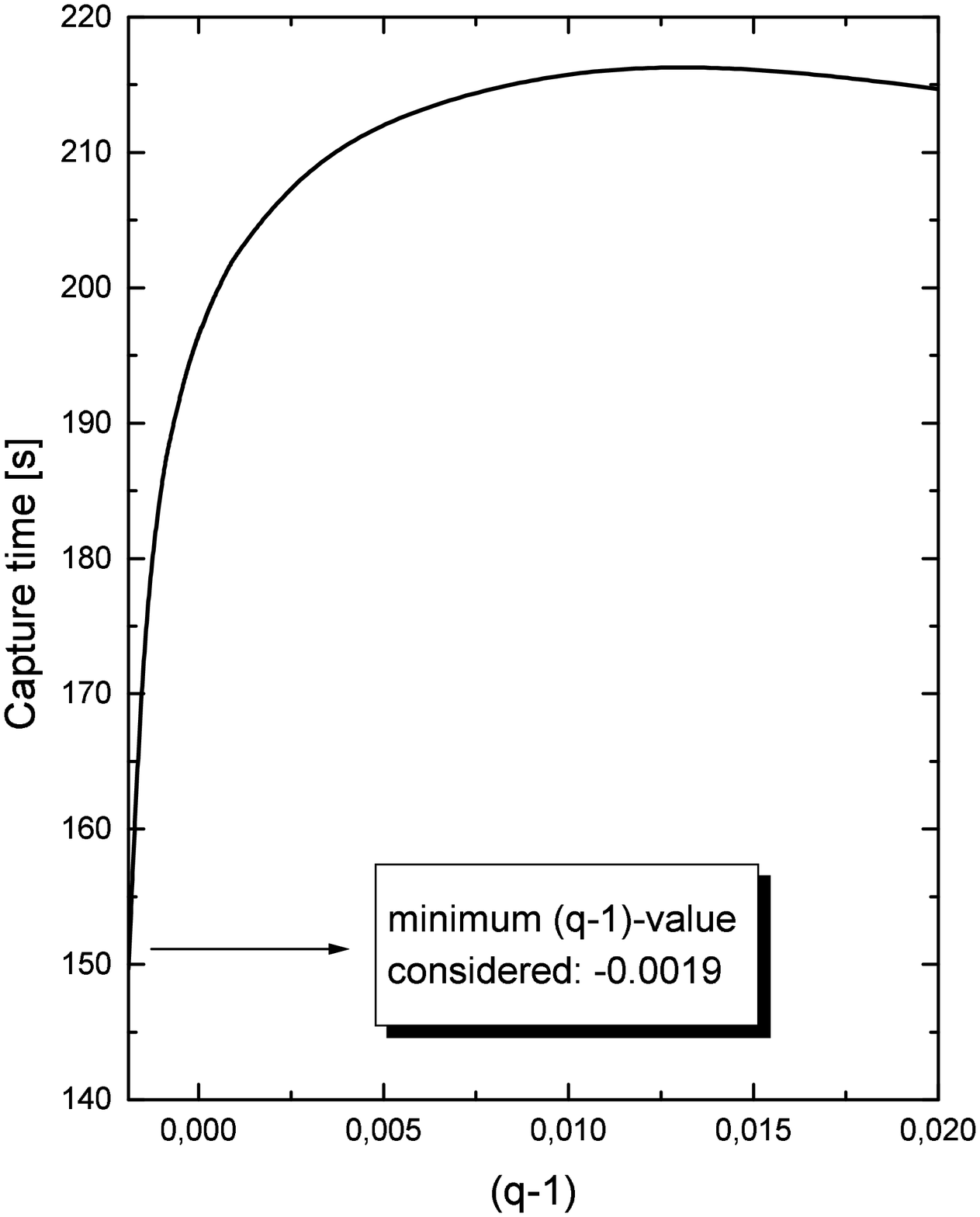}
\end{center}
\vspace{-1.5cm} \caption{Left: $z_c$ as a function of $(q-1)$.
Right: Capture time as a function of $(q-1)$. Both plots are
constructed for a nominal value of $\eta_q=3.57 \times 10^{-10}$.
Note the asymmetric range of the $(q-1)$-values considered. See
text for explanation.} \label{fig:zc}
\end{figure}

Neutron and proton populations are mainly determined by the
kinetic equations
\begin{equation}
\label{eq:dXqn/dz} \frac{dX_q^n}{dz}=-R^q_{np}[X_q^n
X_q^p-G^q_{np}X_q^D] , \hspace{2.3cm} 
\frac{dX_q^p}{dz}=-R^q_{np}[X_q^n X_q^p-G^q_{np}X_q^D].
\end{equation}
As in the usual case, if the neutron population is not depleted by
other reactions such as Eqs. (\ref{eq:reacD2}-\ref{eq:reacD3}),
protons, neutrons and deuterons are kept in equilibrium,
relationships $X^q_p+X^q_n+2X^q_D=1$ and  $X^q_D=(G_{np}^q)^{-1}
X^q_p X^q_n$ being valid. Since $(G_{np}^q)^{-1}$ is very small,
the deuterium number density will be also small, and we can write
a first approximation as
\begin{equation}
\label{eq:XqD1a}
X_q^{D,(1)}=(G_{np}^q)^{-1}X_q^{n,(0)}X_q^{p,(0)},
\end{equation}
where $X_q^{n,(0)}$ and $X_q^{p,(0)}$ are the unperturbed
populations, that obey $X_q^{n,(0)} + X_q^{p,(0)}=1$. Using this
we now have
\begin{equation}
X^q_p+X^q_n \simeq 1-2 (G_{np}^q)^{-1}X_q^{n,(0)}X_q^{p,(0)}.
\end{equation}
Recalling Eq. (\ref{eq:Gqnp}), one sees that for the $z$-values we
are interested ($z \simeq 30$), the major $z$ dependence of
$(G_{np}^q)^{-1}$ goes as $e^z$. Hence, to first order,
\begin{equation}
\frac{d}{dz}(X^q_p+X^q_n)\simeq -2
(G_{np}^q)^{-1}X_q^{n,(0)}X_q^{p,(0)}.
\end{equation}
Adding Eqs. (\ref{eq:dXqn/dz}) and using Eq. (\ref{eq:XqD1a}) we
find that
\begin{equation}
\label{eq:XqD1b} X_q^{D,(1)}=R^q_{np}[X_q^n X_q^p-G^q_{np}X_q^D].
\end{equation}
When deuterium number density is depleted, its number density will
change according to
\begin{equation}
\frac{dX_q^D}{dz}=R^q_{np}[X_q^n X_q^p-G^q_{np}X_q^D]-
R^q_{DD}[2(X_q^D)^2 - G^q_{DD}X_q^T X^q_p] + \ldots
\end{equation}
where $R_{DD}$ is the scaled rate for the reaction
(\ref{eq:reacD2}) and $G^q_{DD}$ is the Saha factor which gives
the equilibrium value for the ratio $(X^q_D)^2/X_q^T X_q^p$. It
can be shown that, in contrast to $G_{np}$, $G_{DD}$ is always a
small number (see Section III of Ref. \cite{Bernstein}). In fact,
in the standard framework $G_{DD}$ is of order $e^{-60}$. We shall
obviously be safe in considering $G^q_{DD}=G_{DD}$, and therefore
in neglecting it. We can obtain $R^q_{DD}$ inserting the value of
$\langle\sigma_{DD}v\rangle_T$ in  Eq. (\ref{eq:Rq}). The
calculation of $\langle\sigma_{DD}v\rangle_T$ is a non-trivial
task, which involves the use of a phenomenological fit to the
cross section plus a thermal average carried out using Boltzmann
velocity distributions for the deuterons. For the sake of
simplicity, we shall adopt the value of the thermal averaged cross
section used by Bernstein et al. This can be written as
\begin{equation}
\langle\sigma_{DD}v\rangle_T=0.016
\frac{z^{2/3}}{\epsilon_D^{7/2}m_{pl}^{5/6}} e^{-1.44z^{1/3}}.
\end{equation}
Introducing the nominal value of $\eta_0$ as we did it when we
wrote $R^q_{np}$, we can now write for $R^q_{DD}$
\begin{equation}
\label{eq:{RqDD}} R^q_{DD}=1.76 \times 10^{7}
\left(\frac{\eta_q}{\eta_0}\right) z^{-4/3} e^{-1.44z^{1/3}}
[1+0.34(q-1)]
\end{equation}
When the deuterium number density decreases enough, the chain of
reactions that converts almost all neutrons into helium is
initiated. We may therefore identify the temperature
$T_{\gamma,c}$, at which the neutrons are captured, or
equivalently, the value of $z=z_c=\epsilon_D/T_{\gamma,c}$, as in
the standard case, by the condition that
\begin{equation}
\left(\frac{dX_q^D}{dz}\right)_{z=z_c}\simeq 0.
\end{equation}
Once we have neglected the factor $G^q_{DD}$, the condition stated
above, together with Eqs. (\ref{eq:XqD1a}) and (\ref{eq:XqD1b})
and the approximation $X_q^D=X_q^{D,(1)}$, gives
\begin{equation}
2X_q^{D,(1)}R^q_{DD}\simeq 1.
\end{equation}
In order to solve this equation we proceed as follows. Let us
first rewrite the previous equation as
\begin{equation}
G^q_{np}=2X_q^{n,(0)}X_q^{p,(0)}R^q_{DD}.
\end{equation}
Note that we can put the rhs in terms of $z$. In Section 8 we have
already given the asymptotic value of $X_q^{n,(0)}$ as
$0.15+0.23(q-1)$, and therefore the value of $X_q^{p,(0)}$ can be
found trough the relation $X_q^{n,(0)} + X_q^{p,(0)}=1$. Now,
using Eq.(\ref{eq:{RqDD}}) leads, to first order in $(q-1)$, to
\begin{equation}
\label{eq:Gqnpz1} G^q_{np}=4.5\times
10^{5}\left(\frac{\eta_q}{\eta_0}\right)
[1+(q-1)1.6]z^{-4/3}e^{-1.44z^{1/3}}.
\end{equation}
On the other hand, we also have an expression for $G^q_{np}$: it
is given by Eq. (\ref{eq:Gqnp}). Therefore, if we were are able to
write Eq. (\ref{eq:Gqnp}) in terms of $z$ we can equal both
expressions and solve for the unknown $z_c$, which will satisfy
the equality for a given value of $(q-1)$. Unfortunately, this is
not easy. The difficulty is due to the presence of the chemical
potentials in $G^q_{np}$ in combinations other than the relation
$\mu_D=\mu_n+\mu_p$. In this sense, we confront here a problem
similar to that found in the recombination study of Paper I. We
shall just mention that it is possible to show that the function
$u_n+u_p-u_D$ has $10^3$ as an upper bound, see Appendix for
details. Therefore, in the range of $z$-values that concerns us,
the corrections in $G^q_{np}$ due to non-extensivity will be at
most of order $[(q-1)/2] 10^3$ and we can then write for
$G^q_{np}=G^q_{np}(z)$,
\begin{equation}\label{eq:Gqnpz2}
G^q_{np}=2.98 \times 10^{12} \left(\frac{\eta_0}{\eta_q}\right)
[1+5\times 10^2(q-1)]z^{3/2} e^{-z}.
\end{equation}
Thus, equating expressions (\ref{eq:Gqnpz1}) and
(\ref{eq:Gqnpz2}), and taking $\eta_q=\eta_0$, we finally found
that $z_c$ must satisfy the following relation,
\begin{equation} \label{ult}
1=1.5 \times 10^{-6}\frac{1+1.6(q-1)}{1+5\times 10^2(q-1)}
z_c^{-17/6} e^{-1.44z_c^{1/3}} e^{z_c}.
\end{equation}
The standard result (i.e. $(q-1)=0$) gives $z_c=27.08$. The
numerical solution of this equation for different values of the
parameter $(q-1)$, and the corresponding capture time, are shown
in Fig. (\ref{fig:zc}). It is interesting to notice that while
values of $(q-1)>0$ can be used within the analytical study
presented in this Section without any limit, this is not the case
for values of the parameter such that $(q-1)<0$. The problem
arises in that in these cases, $G^q_{np}$ can become negative,
what clearly has no sense. In particular, for $(q-1)=-0.002$, the
denominator in Eq. (\ref{ult}) is identically zero, while for
values of $(q-1)<-0.002$ it is negative. Although this mines the
physical viability of the description for these values of $q$, in
the same sense that what was referred before, other assumptions
are taken in this Section, and only a numerical code can give a
more precise answer in either way. Note that to obtain Fig. 7 we
have made use of the non-extensive time-temperature correction. In
Fig. 8 (left panel), we show the dependence of the $^4$He
abundance as a function of $(q-1)$, taking into the free neutron
decay correction.

Some comments about Figs. 7 and 8 are worth doing. Firstly, note
that the dependence of the $^4$He abundance, when neutron free
decay is taken into account, is no longer linear as it was before,
see for instance Eq. (\ref{eq:Yqp}). This makes the study of the
nucleosynthesis process much more interesting, and the use of the
full numerical code a worth doing task. Now we know that even when
considering a first (linear) order correction in the quantum
distribution functions, the final output in the primordial
nucleosynthesis cannot be reduced to a linear correction in the
abundances. Secondly, it appears that for a given value of $Y_p$
(and $\eta$) there are two values of $(q-1)$ that could fit well
the observations. Finally note that the y-axis of Fig. 8 (left
panel) is showing variations at the level of less than 1\%, and in
principle, all values of $(q-1)$ shown there are admitted by the
current constraints on the primordial abundance of $^4$He.
\begin{figure}[t]
\vspace{-2cm}
\begin{center}
\includegraphics[width=8cm,height=11cm]{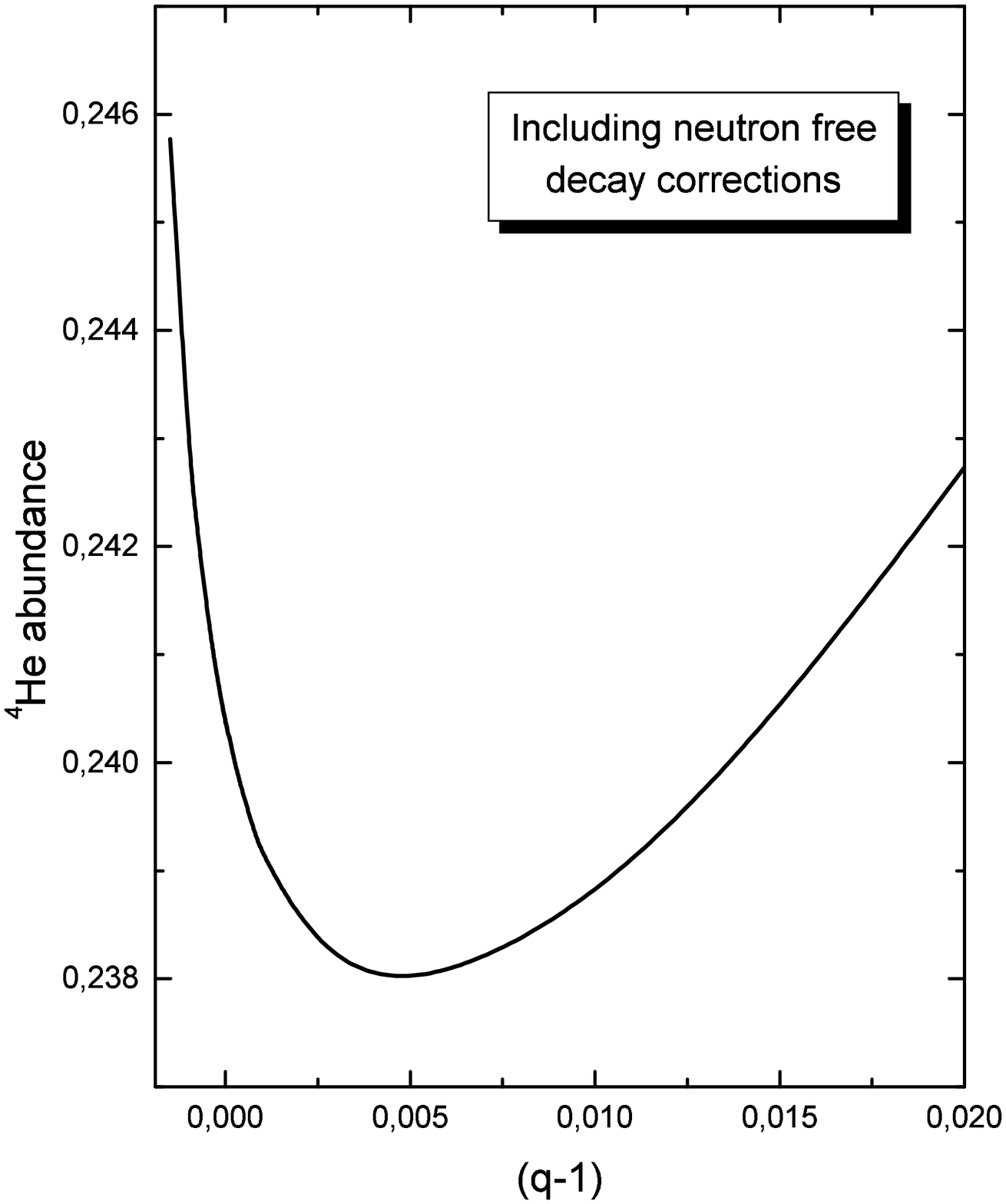}
\includegraphics[width=8cm,height=11cm]{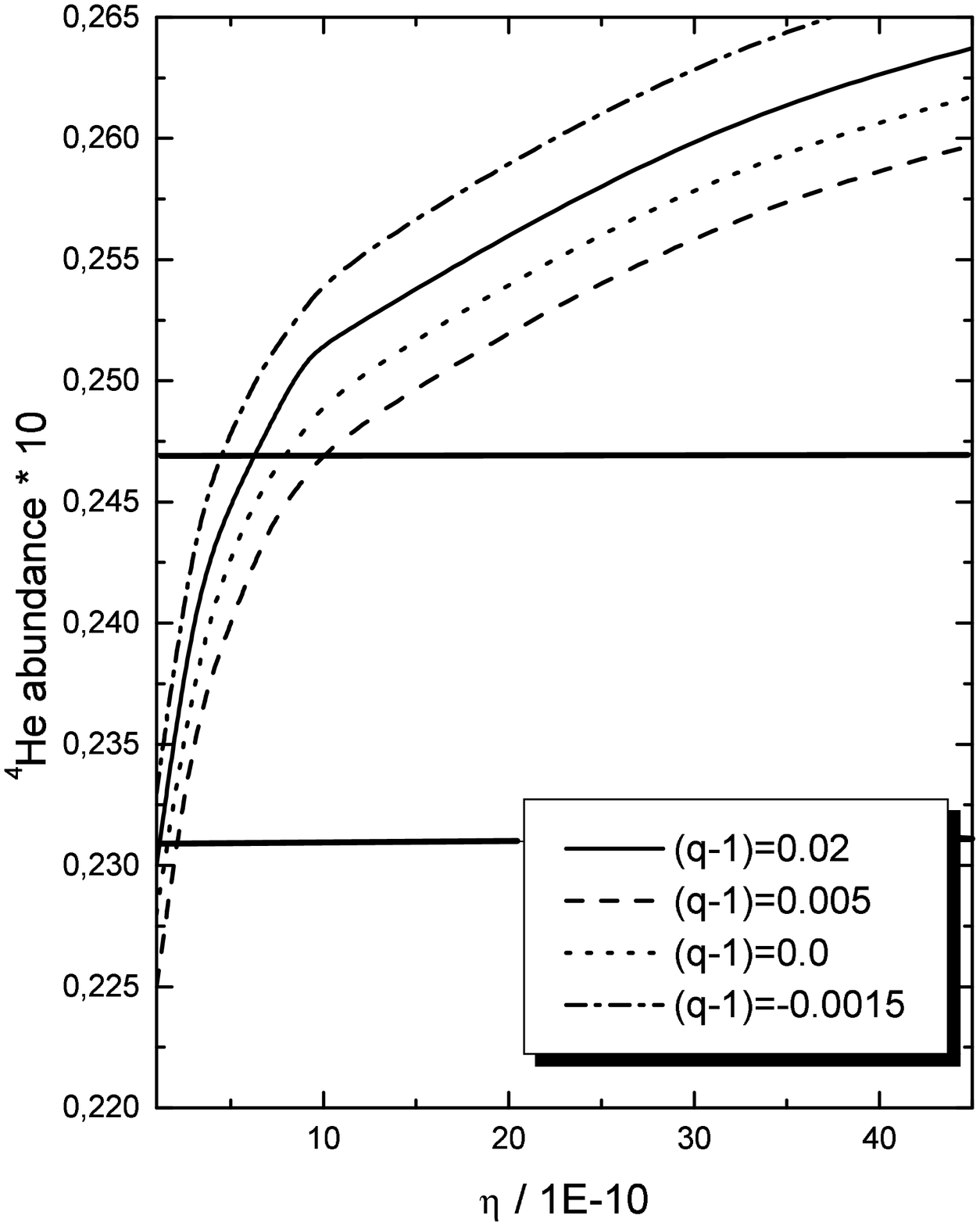}
\end{center}
\vspace{-1.5cm} \caption{Left: $^4$He abundance as a function of
$(q-1)$, taking into account neutron decay corrections. This plot
is constructed for a nominal value of $\eta_q=3.57 \times
10^{-10}$. Right: Primordial abundance of $^4$He as a function of
$\eta_q$. Several $(q-1)$-values are shown.  Two horizontal lines
stand for a conservative current constraint on $Y_p$.}
\label{fig:eta}
\end{figure}

In Fig. 8 (right panel) we show for the first time the dependence
of the $^4$He abundance as a function of $\eta_q$, for different
$(q-1)$-values. The general trend of an increasing abundance as a
function of $\eta$ is maintained, but for different values of $q$,
the curves cross the current constraints in different positions.
This plot actually represents a 3D problem, with a two parameter
space. The influence of $\eta$ appears when numerically solving
the equation for $z_c$. Indeed, the computation of the correction
to the value of $Y_p$ in the full analytical treatment presented
here has to take into account different contributions. A direct
linear dependence on $(q-1)$ is within the time-temperature
relationship (see the first equations in this section, especially
Eq. (\ref{eq:tTrel})) and in Eq. (\ref{eq:Yqp}). In addition, we
can see that if $z_c$ is smaller than in the standard case, then
$T_c$ will be bigger, and $t_c$ will be smaller than their
respective standard counterparts. Thus, $e^{-t/\tau}$ will in turn
be bigger, what happens for instance in the case of a negative
value of $(q-1)$, and this would provide a bigger $Y_p$. However,
the linear correction in Eq. (\ref{eq:Yqp}) is competing with that
positive deviation; this is what finally makes the curve to
deviate from a straight line.

\section{Conclusions}

In this paper, based on previous works (Paper I), we have
revisited the problem of the primordial genesis of light elements
within the framework of non-extensive statistics. We have
particularly paid attention to the form of the principle of
detailed balance that is valid in this new setting. We have shown
that its usual form is no longer valid, but instead, that it is
just an approximation, dubbed here as the BST. The full, new form
of the principle of detailed balance was derived and named as BNE,
and we have studied it in detail too. By doing so, we were able to
disguise the range of validity of previous approaches to the
topic, and to formally see the origin and range of the deviations
between the standard and non-extensive scenarios for the
primordial production of elements. We have also analyzed the
neutron free decay correction to the capture time, and found that
even when considering a first (linear) order correction in the
quantum distribution functions, the final output cannot be reduced
to a linear correction in the abundances, as Fig. 8 explicitly
show. By comparing with the latest observational data, we have
obtained a new bound on the upper limit of the parameter $|q-1|$.
We think that the bound being presented here (referring also to
Figs. 7 and 8) is obtained within a much more detailed basis than
the previous ones, see for instance \cite{PRL,PHYSICA,TIR}, and
then that this bound should be considered as the more reliable of
them all. Indeed, we believe that the only way to improve it is to
go to a direct implementation of the primordial nucleosynthesis
code. By doing so we would have the possibility of studying the
modifications to the abundances of all light elements in a
simultaneous way. This further extension is currently under
analysis.

\subsection*{Appendix: $u_n+u_p-u_D$}

To have a rough idea of the order of magnitude of the function
$u_n+u_p-u_D$, we need to compute $(m_i-\mu_i)/T$ in the range of
temperatures we are interested in, with $i=n, p,$ and $D$.
Firstly, we note that for this purpose it is enough to consider
the standard values of all these quantities. This is because
non-extensive effects here would introduce only second order
corrections in $(q-1)$, that we are disregarding in our
computation. Recalling the relation
$X_i=(g_i/n_B)(m_iT/2\pi)^{3/2}exp((m_i-\mu_i)/T)$, we see that in
order to estimate the values of $(m_i-\mu_i)/T$ we need to know
the abundances $X_i$. Then, to get $u_n+u_p-u_D$, we have made the
following steps:
\begin{enumerate}
\item Using the time-temperature relation we can put $X_n(t)=exp(-\tau/t) X_n(t_{frezee})$
as an explicit function of $T$. Thus $X_n(T)$ follows, as well as
$(m_n-\mu_n)/T$.
\item Considering $X_n(T)$ together with the obtainable function
$X_n X_p/X_D$ (remember that $\mu_n+\mu_p-\mu_D=0$), and imposing
the (approximate) constrain $X_n+X_p+2X_D=1$, we get the values of
$X_p$ and $X_D$. Then, the corresponding values of $(m_p-\mu_p)/T$
and $(m_D-\mu_D)/T$ can be obtained. \end{enumerate}
These results
allow us to construct $u_n+u_p-u_D$ and thus to see that $10^3$
constitutes an upper bound in the range $0.07$MeV $<T<0.1$MeV
(i.e. $22<z<32$).

\subsection*{Acknowledgments}

M. E. Pessah is supported by Fundaci\'on Antorchas. We thank A.
Lavagno, G. Lambiase, P. Quarati, and H. Vucetich for valuable
discussions. D. F. T. was supported by CONICET as well as by funds
granted by Fundaci\'on Antorchas, and acknowledges the hospitality
provided by the Politecnico di Torino, the University of Salerno,
and the ICTP (Italy) during different stages of this research.


\end{document}